\documentclass[%
superscriptaddress,
nofootinbib,
 amsmath,amssymb,
 aps, physrev,
floatfix,
twocolumn,
]{revtex4-2}

\usepackage{graphicx}
\usepackage{dcolumn}
\usepackage{bm}
\usepackage{float}
\usepackage[usenames]{xcolor}  
\usepackage{hyperref}

\begin{document}


\title{\textbf{On the measurability of Wigner time delays at shape resonances \\ in photodetachment of polyatomic anions} 
}%

\author{Jan Dvo\v{r}\'{a}k}
\altaffiliation[]{jdvorak2@lbl.gov, current address: Department of Physics, University of Central Florida, Orlando, FL}
\affiliation{Chemical Sciences Division, Lawrence Berkeley National Laboratory, Berkeley, California 94720, USA}
\affiliation{Department of Chemistry, University of California, Davis, California 95616, USA}

\author{Jakub Benda}
\affiliation{Charles University, Faculty of Mathematics and Physics, Institute of Theoretical Physics, V Hole\v{s}ovi\v{c}k\'{a}ch 2, 180 00 Prague 8, Czech Republic}

\author{Thomas N. Rescigno}
\affiliation{Chemical Sciences Division, Lawrence Berkeley National Laboratory, Berkeley, California 94720, USA}

\author{Cynthia S. Trevisan}
\affiliation{Department of Oceanography and Natural Sciences, Cal Poly Maritime Academy, Vallejo, California 94590, USA}
\affiliation{Department of Physics, California Polytechnic State University, San Luis Obispo, California 93407, USA}

\author{Robert R. Lucchese}
\affiliation{Chemical Sciences Division, Lawrence Berkeley National Laboratory, Berkeley, California 94720, USA}

\author{C. William McCurdy}
\altaffiliation[]{cwmccurdy@ucdavis.edu}
\affiliation{Chemical Sciences Division, Lawrence Berkeley National Laboratory, Berkeley, California 94720, USA}
\affiliation{Department of Chemistry, University of California, Davis, California 95616, USA}

\date{\today}

\begin{abstract}

The energy dependence of the  complex phases of electron continuum wave functions carries information about electron dynamics.  Streaking and attosecond interference experiments (called RABBIT) seek to measure this energy dependence, and therefore, the time delays of ionization.  The long-range Coulomb interaction  dominates in those experiments, and can obscure the low-energy features of the Wigner time delays that are the object of the measurement.  Photodetachment of electrons from negative ions has no long-range Coulomb interaction, and RABBIT and streaking measurements of photodetachment delays have the potential to reveal  time delays of up to one femtosecond in low-energy features.  We predict the results of such experiments on a particularly interesting polyatomic example, the nitrate anion (NO$_3^-$), for both valence and core electron detachment.  We simulate the  experiments in these cases and analyze the underlying physics of measurements on polyatomic anions where many electron partial waves contribute and find that the angular dependence of the measured delays generally differs from the Wigner delays.  However, we demonstrate that measurements performed for ejection directions close to the polarization of the light sources can directly access the Wigner delays that give a time-dependent window on electron-molecule interactions.  A promising experiment involving core photodetachment of NO$_3^-$ with X-rays is proposed. 

\end{abstract}

\maketitle


\section{\label{sec:introduction}Introduction}

The enormous progress in laser technologies in the last decades has provided new tools for studying electron dynamics of atomic, molecular, and even condensed-phase systems on its natural attosecond time scale~\cite{corkum_attosecond_2007, pazourek_attosecond_2015, nisoli_attosecond_2017, agostini_nobel_2024, cruz-rodriguez_quantum_2024, alexander_attosecond_2025}. 
During emission of an electron induced by a photon absorption from such a system, the interaction of the outgoing photoelectron with the other electrons and nuclei becomes imprinted in the phase of the electron's complex-valued  wave function. The change of the phase with respect to the electron's energy can be interpreted as a Wigner time delay (also known as Eisenbud-Wigner-Smith delay) of the electron's wave packet relative to a free wave packet that is not interacting with the residual system~\cite{eisenbud_1948, wigner_lower_1955, smith_lifetime_1960}.

Several experimental methods based on pump-probe techniques have been developed to characterize attosecond pulses and measure such photoionization time delays. The \emph{reconstruction of attosecond beating
by interference of two-photon transitions} (RABBIT) method~\cite{veniard_phase_1996, paul_observation_2001, muller_reconstruction_2002, isinger_photoionization_2017} and 
\emph{kinetic energy} or \emph{angular streaking} methods~\cite{itatani_attosecond_2002, schultze_delay_2010, yakovlev_attosecond_2010, pazourek_attosecond_2015, ortmann_understanding_2024, eckle_attosecond_2008, eckle_attosecond_2008-1, kheifets_ionization_2022, serov_xuv_2023, driver_attosecond_2024, ji_attosecond_2025} combine a high-frequency laser field, which is responsible for the electron emission, with an additional infrared (IR) laser field. The IR field modifies the photoelectron spectrum in a way that allows determination of the time delay with attosecond accuracy.
However, the interaction of the outgoing photoelectron with the IR field introduces an additional phase that produces a further time delay, often called a continuum-continuum delay, which may obscure the intrinsic single-photon (Wigner) time delay that carries the information about the ionization process~\cite{ivanov_how_2011, dahlstrom2013, baykusheva2017}.

In the photoionization of neutral species, the electron-IR interaction in the long-range Coulomb potential of the positively charged residual system significantly affects the time delay, especially at low electron energies below 20~eV. 
To recover the single-photon delay from experiments, the continuum-continuum delay is typically taken into account by using various asymptotic approximations~\cite{ivanov_how_2011, dahlstrom2013, baykusheva2017} or by explicitly modeling the continuum-continuum transitions in time-independent~\cite{dahlstrom_study_2014, hockett_time_2016, baykusheva2017,  benda2021, benda2022} or time-dependent approaches~\cite{haessler_phase-resolved_2009, serov_time_2017, chacon_attosecond_2018, kheifets_rabbitt_2021}.
Despite this complication, the RABBIT and streaking methods have been used to study the photoionization dynamics in many systems of increasing complexity from noble gases~\cite{schultze_delay_2010, feist_time_2014, guenot_measurements_2014, cattaneo_comparison_2016, alexandridi_attosecond_2021, benda_angular_2025}, molecules~\cite{vos_orientation-dependent_2018, huppert_attosecond_2016, cattaneo_comparison_2016, cattaneo_attosecond_2018, kamalov_electron_2020, loriot_high_2020, nandi_attosecond_2020, driver_attosecond_2024, ji_attosecond_2025}, liquids~\cite{jordan_attosecond_2020}, to solids~\cite{cavalieri_attosecond_2007, neppl_direct_2015, tao_direct_2016, kasmi_effective_2017, siek_angular_2017, ambrosio_spatiotemporal_2019}. Furthermore, time delays may arise in molecular photoionization from IR interaction with permanent dipole moments or from IR-induced transitions between states of the residual ion~\cite{benda2022, benda_dipole-laser_2024, delgado_three-path_2025}. 

In photoionization, the Coulomb time delay, which diverges as the photoelectron energy decreases, dominates the total time delay at low energies and can obscure specifically molecular time delay effects.  In contrast, in photodetachment from negative ions there is no residual Coulomb interaction, and that fact may allow direct observation of long single-photon delays and delays due to the effects of electron correlation.  The measurement of photodetachment delays potentially offers a time-dependent view of low-energy electron-molecule scattering and the rich range of phenomena, including long-lived resonances, known to be associated with those collisions.

Despite this advantage,  photodetachment time delays have received much less attention than the photoionization delays due to challenges related to the preparation of the anions in sufficient densities to conduct such experiments. 
There are a few theoretical works focusing on photodetachment time delays, but to our knowledge there have been no measurements of these delays yet.  

Lindroth and Dahlstr\"om~\cite{lindroth_attosecond_2017} showed that, in the absence of the Coulomb potential in the detachment of F$^-$ and Cl$^-$, the continuum-continuum delay effectively vanishes at high photoelectron energies. However the RABBIT delay still deviates from the Wigner delay below 5~eV.  Furthermore, the delay difference is not universal.   That is, unlike in the Coulomb case for noble gases, it depends on the target with which the photoejected electron interacts~\cite{lindroth_attosecond_2017}.
Saha \textit{et al.}~\cite{saha_dominance_2019, saha_wigner_2019}, Banerjee~\textit{et al.}~\cite{banerjee_effects_2020, banerjee_time_2021}, and Kheifets and Bray~\cite{kheifets_time_2021} studied in detail various aspects of Wigner time delays, such as threshold behavior, angular dependence, and relativistic effects (spin-orbit interaction), for Cl$^-$, Br$^-$, I$^-$, Li$^-$, and H$^-$.
In the molecular case, Benda and Ma\v{s}\'{i}n~\cite{benda_dipole-laser_2024} considered detachment from strongly polar anion BeH$^-$ and showed that adding the dipole-laser coupling delay to the Wigner delay fully describes the RABBIT delay above 5~eV. Rescigno \textit{et al.}~\cite{rescigno_attosecond_2024}  recently investigated femtosecond Wigner delays near shape resonances in the core detachment of C$_2^-$ and CN$^-$ in the molecular frame.

Here, to extend photodetachment studies to polyatomic anions and to stimulate experimental efforts, we present our \textit{ab initio} theoretical study of the ultrafast valence and core detachment dynamics in the particularly appealing case of the nitrate anion (NO$_3^-$).   The measurement of delays in photodetachment from polyatomics using RABBIT or streaking techniques will require a clear understanding of the physics that underlies those measurements when many partial waves contribute to the detachment and measurement processes.  Thus, we begin our study with an analysis of the physics of the measurements relevant to these cases.

After valence electron detachment, the nitrate radical (NO$_3$)  shows a complex non-adiabatic vibronic dynamics due to Jahn-Teller and pseudo-Jahn-Teller effects that have been studied both experimentally by Neumark and co-workers~\cite{weaver_examination_1991, babin_high-resolution_2020} and theoretically by Eisfeld and others
\cite{viel_accurate_2021, williams_simulation_2022, mahapatra_effects_2007, mukherjee_beyond_2018}, in recent studies that are part of a larger literature on the subject.  In our work, we do not consider the nuclear motion but focus on the ultrafast detachment dynamics of valence electrons as well as core electrons from nitrogen and oxygen $1s$ orbitals. 

We will show that, in both the laboratory and molecular frames, the angular dependence of the Wigner delay and the delay measured using RABBIT are different, even at energies for which there is no continuum-continuum delay.  This conclusion is general for molecules, and goes beyond the well-known phenomenon of the effective zero in the RABBIT amplitude for ejection at 90$^\circ$ to the polarization direction of the IR radiation~\cite{busto_fanos_2019}.  Nonetheless, we verify that there is a region around ejection along the polarization direction in which the measured RABBIT delays for polyatomic molecules coincide with the Wigner delay at electron energies in excess of about 5 eV in a typical experiment.

We start with a brief overview of theoretical methods in Sec.~\ref{sec:theoretical_methods}. To understand the angular dependence of single-photon (Wigner) and two-photon time delays of NO$_3^-$, we first study the photodetachment of atomic hydrogen and chlorine anions in Secs.~\ref{sec:hydrogen_dynamics} and~\ref{sec:chlorine_dynamics}. We will show that even in the photodetachment case above 5~eV where the continuum-continuum phases are energy-independent, the angular dependence of single- and two-photon time delays are in general not the same due to the interference of the intermediate electron partial waves. However, we will see that for electron emission along an IR polarization direction, the single- and two-photon delays are essentially identical in photodetachment.

After analyzing the simpler atomic cases to make the basic observations, we will focus first on valence detachment of NO$_3^-$ both in the laboratory and molecular frames (Secs.~\ref{sec:one_photon_dynamics}, \ref{sec:two_photon_dynamics_LF}, and~\ref{sec:two_photon_dynamics_MF}), and then, on the detachment of $1s$ core electrons (Sec.~\ref{sec:core_dynamics}). Our calculations will reveal that the valence and core detachment dynamics at electron energies below 25 eV is dominated by shape resonances with time delays up to 1~fs. Our simulations of RABBIT and streaking processes will show that such long resonance delays would be measurable with negligible continuum-continuum delay, that is, the Wigner delay of shape resonances would be directly obtainable from experiments. In Sec.~\ref{sec:discussion}, we will argue that oxygen $1s$ detachment of NO$_3^-$ is an especially attractive candidate for a feasible photodetachment experiment using self-referencing angular streaking measurements if sufficient densities of the negatively charged anions can be produced. Such self-referencing experiments have been recently performed for photoionization of NO and azabenzene molecules at the LCLS X-ray free-electron laser at the SLAC National Accelerator Laboratory~\cite{driver_attosecond_2024, ji_attosecond_2025}. Finally, we will summarize and conclude our work in Sec.~\ref{sec:conclusions}. Further details of averaging over molecular orientations, Wigner delays at threshold, Born approximation for the RABBIT amplitudes, and RABBIT angular dependence for H$^-$ are given in Appendices~\ref{app:lab_frame}, \ref{app:threshold_delay}, \ref {app:Born_RABBIT_amplitudes} and~\ref{app:rabbitt_hydrogen_derivation}, respectively. The Supplemental Material (SM)~\cite{SM} contains additional figures.

\section{\label{sec:theoretical_methods}Theoretical methods}

We start with a brief overview of theoretical methods used to obtain our results, photoelectron spectra and time delays, presented in the following sections. In this work the fixed-nuclei approximation is used. 
We consider all quantities in the molecular frame first and later discuss the transformation to the laboratory frame. Hartree atomic units are used throughout, unless otherwise specified.  

The angle-resolved cross section $\sigma(E,\hat{k})$ for single-photon ionization or detachment molecular process is given in the dipole approximation and the length gauge by the matrix element of the electronic dipole operator $\hat{\varepsilon}\cdot\vec{r}$ between the initial state $\Psi_0$ and final stationary scattering state $\Psi_{f\vec{k}}^{(-)}$ with outgoing photoelectron that carries asymptotic momentum $\vec{k}$~\cite{Lucchese_McKoy_1982}:
\begin{equation}
    \label{eq:xsec}
    \sigma(E, \hat{k}) = \frac{(2\pi)^2\Omega}{c} \vert \langle\Psi_{f\vec{k}}^{(-)}\vert
        \hat{\varepsilon}\cdot \vec{r}\,\vert\Psi_0\rangle\vert^2,
\end{equation}
where $\Omega$ is photon energy, $c$ is the speed of light, and $E=k^2/2$ is asymptotic photoelectron kinetic energy.
Here, the momentum $\vec{k}$ and photon polarization direction $\hat{\varepsilon}$ are defined with respect
to the molecular frame. Expanding the final-state wave function into photoelectron partial waves $\ell$, $m$, the angle-resolved cross section is given by
\begin{equation}
    \sigma(E, \hat{k}) = \frac{(2\pi)^2\Omega}{c} \Big\vert \sum_{\ell,m,\mu} \varepsilon_\mu M^1_{\ell m \mu}(E) Y_{\ell m}(\hat{k})\Big\vert^2,
\end{equation}
\begin{equation}
    M^1_{\ell m\mu} = \langle\Psi_{f k\ell m}^{(-)}\vert\vec{r}_\mu\vert\Psi_0\rangle,
\end{equation}
where $\mu$ runs over $x$, $y$, and $z$ components of the polarization vector $\hat{\varepsilon}$.

The angle-resolved Wigner time delay for such single-photon ionization or detachment process is then given by the energy derivative of the argument of the dipole matrix element:
\begin{eqnarray}
    \label{eq:wigner_delay}
    \tau_W(E, \hat{k})~ &&= \frac{\partial}{\partial E} \arg\,\langle\Psi_{f\vec{k}}^{(-)}\vert
        \hat{\varepsilon}\cdot \vec{r}\,\vert\Psi_0\rangle \\
    &&= \frac{\partial}{\partial E} \arg \sum_{\ell,m,\mu} \varepsilon_\mu M^1_{\ell m\mu}(E) Y_{\ell m}(\hat{k}).
\end{eqnarray}
If multiple degenerate states of the residual ion/molecule contribute, the total Wigner delay~$\tau_W^\mathrm{total}$ is obtained by summing the partial time delays weighted by the partial cross section for photoionization/photodetachment into the states $\alpha$~\cite{baykusheva2017}:
\begin{equation}
    \tau_W^\mathrm{total} = \frac{\sum_\alpha \tau_{W\alpha}(E,\hat{k})\sigma_\alpha(E,\hat{k})}{\sum_\alpha\sigma_\alpha(E,\hat{k})}.
\end{equation}

In this work, we study single-photon valence photodetachment dynamics of H$^-$, Cl$^-$, and NO$_3^-$ using the 
R-matrix method~\cite{burke2007, burke2011, tennyson2010} as implemented in \textsc{UKRmol+} suite of codes~\cite{masin2020_ukrmolp, houfek2024} and in the case of NO$_3^-$ also using the complex Kohn method~\cite{KohnPhotoionization_1993, rlm95, Miyabe_CO2_2009, Rescigno_Douget_Orel_2012, trevisan_2012} and Schwinger variational method~\cite{Gianturco1994, Natalense1999} as implemented in the \textsc{ePolyScat} suite of codes~\cite{Schneider2020}.
Furthermore, we investigate RABBIT time delays using the second order of the perturbation theory as implemented in the latest version of \textsc{UKRmol+}~\cite{benda2021, benda2022}, where the continuum-continuum transitions are explicitly taken into account. Additionally, we simulate  RABBIT and streaking processes using the R-matrix with time-dependence method~\cite{brown2020_rmt, moore2011}.

In the RABBIT process, two quantum pathways lead to the same final continuum state $\Psi_{f\vec{k}}^{(-)}$: i) 
\emph{absorption path} where absorption of a photon with energy $\Omega_+$ is followed by absorption of an IR photon with energy $\omega$; ii) \emph{emission path} where absorption of a photon with energy $\Omega_- = \Omega_+ + 2\omega$ is followed by emission of one IR photon. 
In the second order of perturbation theory, the dipole matrix elements for the absorption $M^{+}$ and emission 
$M^{-}$ two-photon pathways can be calculated as~\cite{cohen-tannoudji_atom-photon_2008, dahlstrom2013, benda2021, benda2022}
\begin{equation}
    \label{eq:two_photon_element}
    M^{\pm} = -i\langle\Psi_{f\vec{k}}^{(-)}\vert \hat{\varepsilon}\cdot \vec{r}\,\, G^{(+)}(E_\pm)\, \hat{\varepsilon}\cdot \vec{r}\,\vert\Psi_0\rangle,
\end{equation}
where $G^{(+)}(E_\pm)=(E_0+\Omega_\pm - H + i0)^{-1}$ is Green's function for the full Hamiltonian~$H$ of the system and $E_0$ is the energy of the initial state. 
The same linear polarization $\hat{\varepsilon}$ is here assumed for the high-frequency and IR photons.

The interference of the absorption and emission pathways, which have the same final energy $E_0 + \Omega_+ + \omega$, gives rise to oscillations of the RABBIT signal $P(\tau)$ as a function of the time delay $\tau$ between the high-frequency and IR fields~\cite{dahlstrom2013}:
\begin{eqnarray}
    P(t)\, &&\propto \vert M^{+} e^{i\omega\tau} + M^{-} e^{-i\omega\tau}\vert^2 \label{eq:rabitt_signal}\\
      &&= a + b \cos[2\omega(\tau + \tau_R)]\label{eq:rabitt_signal_cos},
\end{eqnarray}
where $a = \vert M^{+}\vert^2 + \vert M^{-}\vert^2$, $b = 2\mathrm{Re}\,M^{+*}M^{-}$, and $\tau_R$ is the RABBIT delay 
\begin{equation}
\label{eq:rabitt_delay}
    \tau_R(E,\hat{k}) = \frac{1}{2\omega} \arg M^{+*}M^{-}
\end{equation}
of the outgoing photoelectron with energy $E$ and direction $\hat{k}$ defined with respect to the molecular frame.
Note that the matrix elements in Eq.~\eqref{eq:two_photon_element} do not include the complex amplitudes for the electric fields. To obtain the \emph{cosine} dependence of the RABBIT signal, the relative field factors  $\exp(\pm i\omega\tau)$ are added in Eq.~\eqref{eq:rabitt_signal} for the IR absorption and emission.

In the RABBIT method, the energy derivative of the phase of the two-photon matrix element is approximated by the finite difference in Eq.~\eqref{eq:rabitt_delay}. Sometimes it is beneficial to compare the RABBIT delay to the Wigner delay approximated in the same way to better judge which effects originate from the IR interaction. The finite-difference approximation to the Wigner delay [Eq.~\eqref{eq:wigner_delay}], sometimes called ``one-photon delay''~$\tau_1$~\cite{benda2022}, is obtained by evaluating the single-photon matrix elements at the harmonic photon energies $\Omega_+$ and $\Omega_-$:
\begin{equation}
    \label{eq:1hv_delay}
    \tau_1(E,\hat{k}) = \frac{1}{2\omega} \arg \sum_{\ell,m,\mu} d^{1}_{\ell m\mu}(\Omega_+)^*\,
    d^{1}_{lm\mu}(\Omega_-),
\end{equation}
where $d^{1}_{\ell m\mu} = \varepsilon_\mu M^1_{\ell m\mu} Y_{\ell m}(\hat{k})$.  Hereafter we follow that convention, and use ``one-photon delay'' to mean this finite-difference approximation to the single-photon Wigner delay in RABBIT measurements.

Finally, we also solve the time-dependent Schr\"{o}dinger equation (TDSE) 
\begin{equation}
    \label{eq:TDSE}
    i\frac{\partial}{\partial t}\Psi(t) = \left[ H + \vec{\mathcal{E}}(t)\cdot\vec{r}\,\right]\Psi(t)
\end{equation}
using the R-matrix with time-dependence (RMT) method~\cite{moore2011, brown2020_rmt} to simulate the RABBIT and kinetic energy streaking processes. Here, $\Psi(t)$ and $H$ represent the total electronic wave function and Hamiltonian of the system, respectively.

For the simulation of the RABBIT process, $\vec{\mathcal{E}}(t)$ in Eq.~\eqref{eq:TDSE} denotes the sum of electric fields for
the IR pulse and attosecond pulse train, which has a spectrum consisting of odd harmonics with energies $(2n+1)\omega$~\cite{serov_time_2017, agostini_nobel_2024}. 
The TDSE is then solved for a set of time delays~$\tau$ between the attosecond pulse train and the IR field to obtain the photoelectron spectrum $P(E,\hat{k}, \tau)$. This spectrum manifests quantum beating with frequency $2\omega$ as in Eq.~\eqref{eq:rabitt_signal_cos} at the RABBIT sideband electron energies $E_n=2n\omega-E_\mathrm{IP}$, where $E_\mathrm{IP}$ is the ionization potential of the system.
By fitting Eq.~\eqref{eq:rabitt_signal_cos} to the spectrum at $E_n$ with $a$, $b$, and $\tau_R$ as fitting  parameters, the RABBIT time delay $\tau_R(E_n,\hat{k})$ can be extracted. 

In the streaking process, $\vec{\mathcal{E}}(t)$ in Eq.~\eqref{eq:TDSE} describes a single attosecond pulse combined with the IR field. When the attosecond 
pulse ionizes or detaches an electron that would have asymptotic momentum~$\vec{k}_0$, the electron is ``born" in the continuum with additional momentum given by the vector potential $\vec{A}_\mathrm{IR}$ of the IR field~\cite{pazourek_attosecond_2015}:
\begin{equation}
    \label{eq:streaking}
    \vec{k}_f(\tau) = \vec{k}_0 - \vec{A}_\mathrm{IR}(\tau + \tau_S).
\end{equation}
The effect of the IR field is delayed by the streaking time delay~$\tau_S$.
We can use Eq.~\eqref{eq:streaking} with the known form of the vector potential to fit the calculated photoelectron spectrum as a function of delay~$\tau$ between the attosecond and IR pulses in order to extract the streaking delay $\tau_S(E, \hat{k})$.

Up to this point, the direction of the outgoing photoelectron~$\hat{k}$ and the field polarizations~$\hat{\varepsilon}$ were defined with respect to the molecular frame, and thus, the cross sections (spectra) and time delays depend on the molecular orientation. In principle, the molecular frame is accessible in the experiments with oriented molecules~\cite{stapelfeldt_colloquium_2003, seideman_nonadiabatic_2005, wang_molecular_2025} or in electron-ion coincidence measurements assuming axial recoil approximation~\cite{dorner_cold_2000, schmidtbocking_coltrims_2021}. However, aligning anions with nonresonant laser pulses or identifying cases with rapid dissociation following the photodetachment allows coincidence measurements to detect molecular orientation, significantly increases the complexity of the potential experiments to measure molecular-frame photodetachment time delays.

Measurements in the laboratory frame are simpler, and therefore, we discuss mostly time delays calculated in the laboratory frame in this work. To transform to the laboratory frame, we numerically average the Wigner, one-photon, and perturbation-theory RABBIT delays over molecular orientations assuming randomly oriented molecules. Details can be found in Appendix~\ref{app:lab_frame}.    We do not present our TDSE results averaged over molecular orientations  because doing so requires solving the TDSE for many polarization directions separately.   That calculation would be prohibitive because solving the TDSE for a single polarization direction is computationally quite costly.


In our calculations presented in the following sections, the total final wave function $\Psi_{f\vec{k}}^{(-)}$ of the electron-atom/molecule complex with $N+1$ electrons is represented by the close-coupling expansion
\begin{equation}
    \label{eq:close_coupling}
    \Psi_{f\vec{k}}^{(-)}(\mathbf{r}^{N+1}) = \mathcal{A}\sum_\alpha \Phi_\alpha(\mathbf{r}^N)\chi_{\alpha,\vec{k}}(\mathbf{r}),
\end{equation}
where $\Phi_\alpha$ are $N$-electron wave functions, so-called target channels, that describe states of the neutral atom/molecule after the photodetachment, 
$\chi_{\alpha,\vec{k}}$ are single-electron continuum wave functions for the outgoing electron in the target channels, and
$\mathcal{A}$ is the antisymmetrizing operator. Orbitals for bound electrons in the target states are calculated using quantum chemistry methods, while the continuum wave functions $\chi_{\alpha,\vec{k}}$ are obtained from scattering calculations with R-matrix, complex Kohn, or Schwinger approaches, some details of which are given in the sections below.

\section{\label{sec:hydrogen_dynamics}One- and Two-Photon Detachment Dynamics of H\boldmath{$^-$}}

\begin{figure*}[t!]
\includegraphics[width=1.0\textwidth]{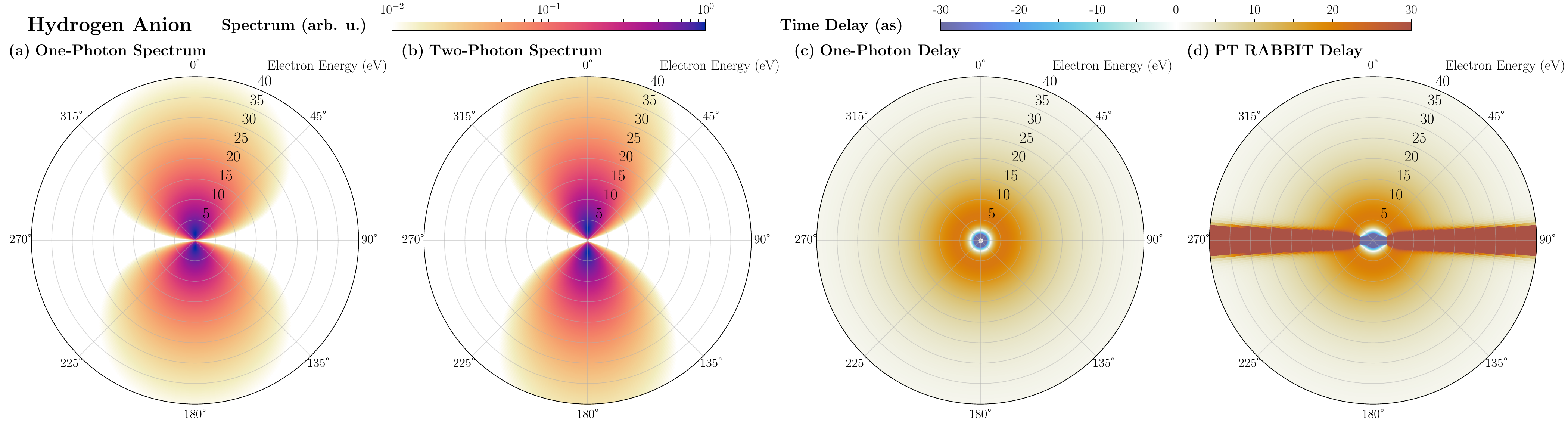}
\caption{\label{fig:polar_plots_hydrogen}
One- and two-photon detachment of atomic hydrogen anion in the static-exchange approximation. Panels~a and~b show the one- and two-photon spectra in a logarithmic scale, respectively. Panels~c and~d show the one-photon and perturbation-theory RABBIT time delays (with IR energy of 0.517~eV) in a linear scale, respectively. The field polarization is along the vertical axis ($z$-axis).
}
\end{figure*}

\begin{figure}[]
\includegraphics[scale=1]{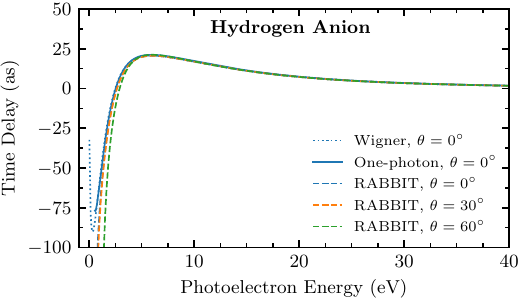}
\caption{\label{fig:delays_hydrogen_1d}
Wigner, one-photon, and perturbation-theory RABBIT time delays for photodetachment of atomic hydrogen anion for several angles between the polarization and electron emission directions. The Wigner and one-photon delays are angle independent.
}
\end{figure}

\begin{figure*}[t!]
\includegraphics[width=1.0\textwidth]{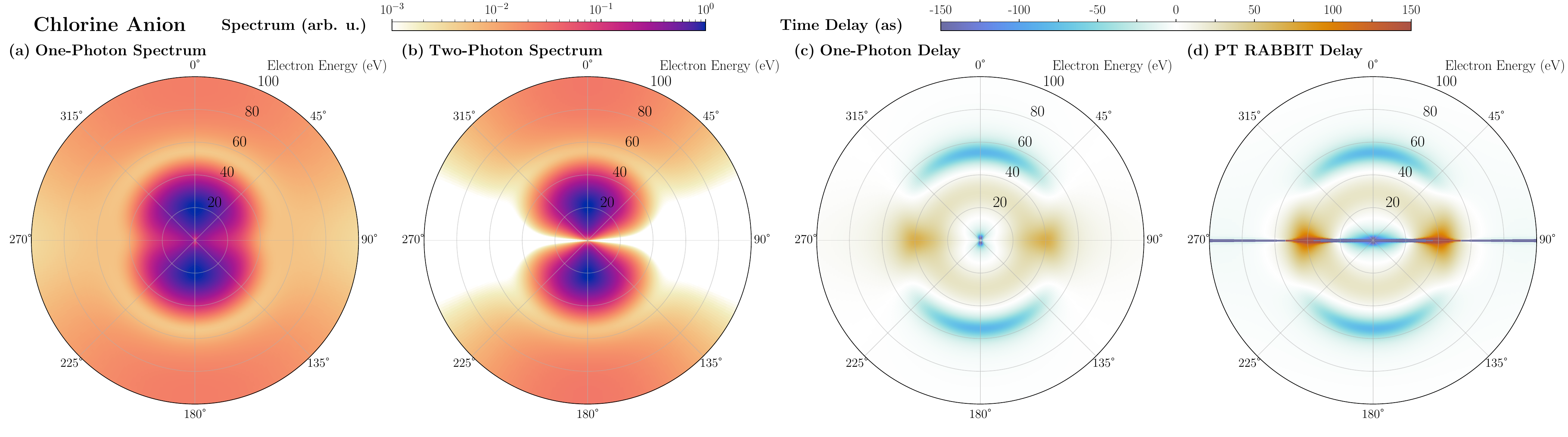}
\caption{\label{fig:polar_plots_chlorine}
One- and two-photon detachment of atomic chlorine anion from the $3p$ orbital in the static-exchange approximation. Panels~a and~b show the one- and two-photon spectra in a logarithmic scale, respectively. Panels~c and~d show the one-photon and perturbation-theory RABBIT time delays (with IR energy of 0.517~eV) in a linear scale, respectively. The field polarization is along the vertical axis ($z$-axis).
}
\end{figure*}

\begin{figure}[]
\includegraphics[scale=1]{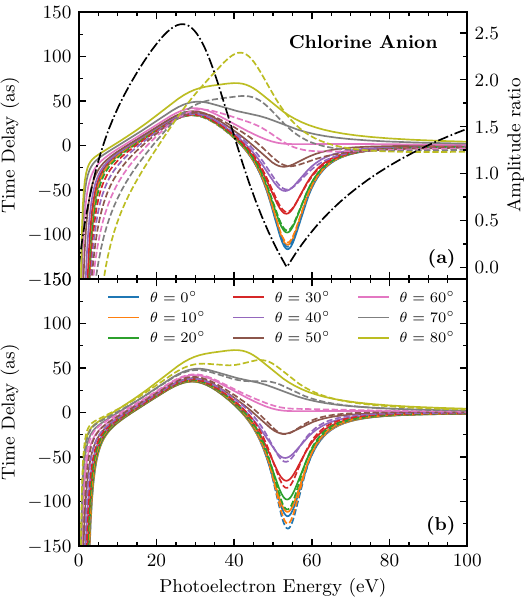}
\caption{\label{fig:delays_chlorine_1d}
Angular dependence of time delays for photodetachment of atomic chlorine anion from the $3p$ orbital. Panel~a shows lineouts of the one-photon (solid lines) and perturbation-theory RABBIT (dashed) delays for several angles~$\theta$. The ratio $\vert M^1_{p\to d_0}\vert / \vert M^1_{p\to s}\vert$ of one-photon dipole matrix elements for the final $d_0$ and $s$ partial waves is plotted as the dash-dotted line with respect to the right-hand vertical axis. Panel~b compares the one-photon delay with RABBIT delays where the continuum-continuum contributions were eliminated, see the text. 
}
\end{figure}

We start the discussion of our results with the simplest photodetachment case: the photodetachment of atomic hydrogen anion~H$^-$ in its ground electronic state $^1S\,(1s^2)$. Upon absorbing a single photon (of sufficient energy) polarized along the $z$-axis, the $1s$ electron is detached as a pure $p_0$ wave ($\ell=1$, $m=0$) and the residual neutral hydrogen atom is left in its ground state~$^2S\,(1s^1)$. As a result, the single photoelectron angular distribution has a $\cos^2\theta$ angular shape:
\begin{equation}
    P_1(E,\theta) = \vert M^1_{s\to p_0}(E)\vert^2\, \vert Y_{10}(\theta)\vert^2,
\end{equation}
where $\theta$ is the polar angle between the polarization axis ($z$-axis) and electron emission direction, $M^1_{s\rightarrow p_0}(E) = \langle \Psi^{(-)}_{f k p_0}\vert z\vert\Psi_0\rangle$ is the single-photon dipole matrix element for the final electron energy $E=k^2/2$,
and $Y_{10}$ is the corresponding spherical harmonic function.
The spectrum is isotropic in the azimuthal angle~$\varphi$ for linearly polarized light. Here and in what follows we suppress the dependence of the spherical harmonics on~$\varphi$. 

The one-photon time delay~$\tau_1$, by which we mean the finite-difference approximation to the Wigner delay  in Eq.~\eqref{eq:1hv_delay}, to be compared with the RABBIT delay, is for the hydrogen anion given by 
$\tau_1(E) = \delta\varphi_1(E)/(2\omega)$ with $\delta\varphi_1(E) = \eta_p(E+\omega) - \eta_p(E - \omega)$.  Here, $\eta_p$ is the phase of the one-photon matrix element $M^1_{s\rightarrow p_0}(E) = \vert M^1_{s\to p_0}(E)\vert e^{i\eta_p(E)}$ and  
$E-\omega$ and $E+\omega$ are the harmonic energies for absorption and emission paths in the RABBIT process.
In this case, where a single partial wave contributes, the one-photon delay is isotropic.

Using the \textsc{UKRmol+} suite of codes, we performed the described calculation for the single-photon detachment of H$^-$ in the static-exchange (SE) approximation. To represent the $^2S$ target state, we used the $1s$ Hartree-Fock (HF) orbital optimized for the anion using \textsc{MOLPRO}~\cite{molpro_2012, molpro_2020, molpro_version_2022d3} with Dunning augmented correlation-consistent triple-zeta (aug-cc-pVTZ) Gaussian basis set~\cite{dunning1989}. Using an R-matrix box of 120~bohr, the continuum electron was represented with 240 B-splines of ninth order and maximum angular momentum $\ell_\mathrm{max}=5$. 

The resulting one-photon spectrum and time delay are shown in Figs.~\ref{fig:polar_plots_hydrogen}a and~\ref{fig:polar_plots_hydrogen}c. 
In the SE calculation, the one-photon delay has a maximum of 21~as at 6~eV and goes to zero at higher energies, see also Fig.~\ref{fig:delays_hydrogen_1d}. Because no polarizability of the residual hydrogen atom is included in this calculation, the Wigner delay goes to zero at 
zero energy, a result given by effective range theory for $p$-wave electron-atom scattering. Modifying the effective range formula for the polarization potential leads to finite delay of 46~as at the threshold for hydrogen anion, as discussed in  Appendix~\ref{app:threshold_delay}.

In the RABBIT process for H$^-$, the absorption or emission of an additional $z$-polarized IR photon leads to final $s$ ($\ell=0$, $m=0$) and $d_0$ ($\ell=2$, $m=0$) electron partial waves. The two-photon RABBIT angular distribution that also depends on the time delay~$\tau$ between the high-frequency and IR fields has the following form:
\begin{align}
    \label{eq:2hv_spectrum_hm}
    P_2(E,\theta,\tau) &= \vert A_0(E,\tau) Y_{00}(\theta) + A_2(E,\tau) Y_{20}(\theta)\vert^2,\\
    A_0(E,\tau) &= M^+_{p \to s}\, e^{i\omega\tau} + M^-_{p \to s}\, e^{-i\omega\tau}, \\
    A_2(E,\tau) &= M^+_{p \to d}\, e^{i\omega\tau} + M^-_{p \to d}\, e^{-i\omega\tau}
\end{align}
where $M^\pm_{\ell\to\ell_f}$ are the two-photon dipole matrix elements [Eq.~\eqref{eq:two_photon_element}] from intermediate $\ell$ to final $\ell_f$ wave (we suppress the $m$ quantum numbers).

Assuming that the phases of the two-photon matrix elements in the short-range potential can 
be written as sum of the one-photon phase $\eta_p^\pm$, evaluated at the harmonic energies, $\eta^\pm_p=\eta_p(E\mp\omega)$, and the continuum-continuum phases, $\varphi^\pm_{s/d}=\varphi_{s/d}(E\mp\omega)$, we can write the amplitudes as,
\begin{align}
  M^\pm_{p \to s} &= |M^\pm_{p \to s}|\, e^{i(\eta_p^\pm + \varphi_s^\pm)}, \\
  M^\pm_{p \to d} &= |M^\pm_{p \to d}|\, e^{i(\eta_p^\pm + \varphi_d^\pm)},
\end{align}
which is supported by the analytical calculation of the two-photon amplitudes in the Born approximation presented in Appendix~\ref{app:Born_RABBIT_amplitudes}.

The square in Eq.~\eqref{eq:2hv_spectrum_hm} leads to four oscillating factors $c_i(\theta)\cos(2\omega\tau+\Delta\Phi_i)$,
with different phase offsets $\Delta\Phi_i$:
\begin{align}
 \Delta\Phi_0 &= \delta\varphi_1 + \varphi_s^- - \varphi_s^+, \\
 \Delta\Phi_2 &= \delta\varphi_1 + \varphi_d^- - \varphi_d^+ \\
 \Delta\Phi_{02a} &= \delta\varphi_1 + \varphi_d^- - \varphi_s^+, \\
 \Delta\Phi_{02b} &= \delta\varphi_1 + \varphi_s^- - \varphi_d^+, 
\end{align}
as shown in Appendix~\ref{app:rabbitt_hydrogen_derivation}.

The Born approximation results in Appendix \ref{app:Born_RABBIT_amplitudes}  show that in  photodetachment at sufficiently high electron energies $\varphi^+_{\ell_f}=\varphi^-_{\ell_f}= -\pi/2$ and
$\varphi^\pm_{\ell_f}=\varphi^\pm_{\ell_f^{'}}= -\pi/2$, and thus, all the individual phase offsets above reduce to the one-photon phase offset $\delta\varphi_1$. As a result, the angular dependence can be factored out
\begin{equation}
    \label{eq:hm_rabbitt_angle_indep}
  P_2(E,\theta, \tau)
  = G(\theta) \cos(2\omega\tau + \delta\varphi_1)
\end{equation}
and the RABBIT delay is angle-independent and equal to the one-photon delay.

In addition to the one-photon dynamics, we also numerically calculated the RABBIT spectrum and time delays for photodetachment of H$^-$ within the second-order perturbation theory using the R-matrix framework [Sec.~\ref{sec:theoretical_methods}], see Figs.~\ref{fig:polar_plots_hydrogen}b and~\ref{fig:polar_plots_hydrogen}d. We emphasize that the R-matrix approach numerically calculates the full two-photon matrix elements for the absorption and emission pathways without relying on any asymptotic approximations for the continuum-continuum transitions, and thus is valid for arbitrary electron energies~\cite{benda2021, benda2022}. The numerical calculations confirm that the continuum-continuum phases are basically the same as the Born approximation limit and energy-independent above 5--10~eV (Appendix~\ref{app:Born_RABBIT_amplitudes}).

Even though the final partial waves are $s$ and $d_0$ for the RABBIT process, the dynamics is dominated by the one-photon transition to the same intermediate $p_0$ wave, and as a result, the RABBIT signal basically vanishes at angles near $\theta=90^\circ$ (Fig.~\ref{fig:polar_plots_hydrogen}b). More specifically, Busto \textit{et al.}~\cite{busto_fanos_2019} showed for the photoionization of helium that due to the asymmetry between the absorption and emission paths, one of the corresponding two-photon matrix elements goes through zero close to $\theta=90^\circ$, which leads to a $\pi$~jump of the RABBIT phase, and consequently, to a $\pi/(2\omega)$ jump in the RABBIT delay (see Fig.~3 in Ref.~\cite{busto_fanos_2019}). The photodetachment of H$^-$ is symmetry-wise equivalent to photoionization of helium and we observe the same effect in the RABBIT delay close to $\theta=90^\circ$, as seen in  the narrow sector around $\theta=90^\circ$ in Fig.~\ref{fig:polar_plots_hydrogen}d. The sign of the $\pi/(2\omega)$ jump, which is 2~fs for $\omega=0.517$~eV, depends on electron energy and changes when the delay changes sign around 3~eV (Fig.~\ref{fig:delays_hydrogen_1d}), which results in a $2\pi/(2\omega)$ discontinuity around 3~eV and $\theta=90^\circ$ visible in Fig.~\ref{fig:polar_plots_hydrogen}d as the sharp change from blue to reddish color. 

As indicated by Eq.~\eqref{eq:hm_rabbitt_angle_indep} and discussed in Appendix~\ref{app:rabbitt_hydrogen_derivation}, the RABBIT angular dependence for H$^-$ is weak and restricted to low electron energies below 5~eV as is visible in Fig.~\ref{fig:delays_hydrogen_1d}. At these low energies the short-range potential in the photodetachment case (exponentially decaying in the SE approximation) affects the continuum-continuum transitions and the $\Delta\Phi_i$ offsets are not exactly equal to the one-photon phase offset. 

 Photodetachment  from the hydrogen anion leads to angle-independent RABBIT delay due to a single intermediate partial wave and cancellation of continuum-continuum phase differences at energies above 5~eV. This is the simplest case for the relation of the measured RABBIT delay to  the Wigner delay.  Many other atomic cases and almost all molecular photodetachment cases differ significantly.

\section{\label{sec:chlorine_dynamics}One- and Two-Photon Detachment Dynamics of C\lowercase{l}\boldmath{$^-$}}

Next, we examine the more complicated case of photodetachment of atomic chlorine anion Cl$^-$ in its ground state $^1S\,(3p^6)$.
Detaching one $3p$ electron by a single $z$-polarized photon results in $s$ and $d_0$ electron partial waves with residual chlorine atom in the $^2P^o\,(m_t=0)$ state and additional $d_{\pm1}$ waves for the $^2P^o\,(m_t=\mp 1)$ residual atom.

In the chlorine case, the single-photon spectrum is given by the incoherent sum over the three target states with $m_t=0,+1,-1$:
\begin{equation}
    P_1(E,\theta) = \vert M^1_{s} Y_{00}(\theta) + M^1_{d_0}Y_{20}(\theta)\vert^2 
    + 2 \vert M^1_{d_1}\vert^2 \vert Y_{21}(\theta)\vert^2,
\end{equation}
where $m_t=0$ has the coherent contribution of the two partial waves and the other residual ion target channels  $m_t=\pm1$ give the same result.

We can write the single-photon matrix elements as 
\begin{equation}
    M^1_{\ell_f m_f}(E) = c_{\ell_f m_f} \vert R^1_{\ell_f}(E)\vert e^{i\eta_{\ell_f}(E)},
\end{equation}
where $\eta_{\ell_f}(E)$ is the energy-dependent phase of the radial part of the dipole matrix element $R^1_{\ell_f}(E)$, which is independent of the $m_f$ quantum number, and $c_{\ell_f m_f}$ is the remaining angular part, which may have different signs for different $m_f$ values.
For the $m_t=\pm 1$ channels, only one partial wave ($d_{\pm 1}$) contributes and the single-photon phase is given by $\arg [M^1_{d_1}(E)Y_{21}(\theta)]$. The additional phase factor from the angular part $c_{\ell_f m_f}Y_{21}(\theta)$ is energy independent and when the finite-difference is taken, these factors cancel each other and the one-photon phase difference is 
$\delta\varphi_1^{(m_t=\pm1)} = \eta_d^- - \eta_d^+$ for the $m_t=\pm 1$ channels, where $\eta_d^\pm = \eta_d(E \mp\omega)$.

In the $m_t=0$ channel, $s$ and $d_0$ waves interfere and the relevant phase is
\begin{equation}
\varphi_1^{m_t=0}(E,\theta) = \arg\!\Big[M^1_s(E)Y_{00} + M^1_{d_0}(E)Y_{20}(\theta)\Big].
\end{equation}
Factoring out the $d$-wave phase, we obtain
\begin{align}
\label{eq:clm_1hv_delay}
\varphi_1^{m_t=0}&(E,\theta) = \eta_d(E)\nonumber \\
&+ \arctan\left(\frac{|M^1_s|\,Y_{00}\,\sin\Delta\eta(E)}{|M^1_d|\,Y_{20}(\theta) + |M^1_s|\,Y_{00}\,\cos\Delta\eta(E)}\right),
\end{align}
where $\Delta\eta(E) = \eta_s(E) - \eta_d(E)$. The one-photon phase difference is then $\delta\varphi_1^{m_t=0}(E,\theta)=\varphi_1^{m_t=0}(E+\omega,\theta)-\varphi_1^{m_t=0}(E-\omega,\theta)$, which has a nontrivial angular dependence because of the $Y_{20}(\theta)$ in the $\arctan$ formula.

We performed R-matrix calculations, analogous to our calculations on $H^-$, of the single-photon and RABBIT dynamics for photodetachment of Cl$^-$ in the SE approximation. 
The bound electrons were represented using state-averaged HF orbitals for $^2P^o_x$, $^2P^o_y$, and $^2P^o_z$ states of Cl, calculated using \textsc{MOLPRO}~\cite{molpro_2012, molpro_2020, molpro_version_2022d3} with cc-pVTZ basis~\cite{dunning1989}.
The continuum electron was represented in the R-matrix sphere of 120~bohr with 240 B-splines of ninth order and maximum angular momentum $\ell_\mathrm{max}=6$. 
The resulting one-photon angle-resolved spectrum and time delay are shown in Figs.~\ref{fig:polar_plots_chlorine}a and~\ref{fig:polar_plots_chlorine}c. Additionally, lineouts of the one-photon delay are shown by solid lines in Fig.~\ref{fig:delays_chlorine_1d}.

The single-photon dynamics is dominated by the Cooper minimum~\cite{cooper_photoionization_1962, cukras_coupled-cluster_2014, lindroth_attosecond_2017} at 54~eV in our SE calculation, where the $d$-wave matrix elements go through zero, and the interference with the $s$~wave results in a negative delay of 100~as for $\theta=0^\circ$ (Fig.~\ref{fig:delays_chlorine_1d}a). Even though the Wigner/one-photon phase is angle-independent at the Cooper minimum (as can be seen from Eq.~\eqref{eq:clm_1hv_delay}), the delay is given by the energy 
derivative of the phase, which for the $d$-wave significantly changes and gives rise to the strong angular dependence in this energy region.
At other energies the $d$~wave dominates (see the dash-dotted curve in Fig.~\ref{fig:delays_chlorine_1d}a) and the spectrum weakly depends on the emission angle. The $s$ wave dominates at  threshold and the delay diverges.  That divergence can be seen in  the analytic low-energy behavior of electron-atom scattering phase shift,  discussed in Appendix~\ref{app:threshold_delay}. 

In the RABBIT process for Cl$^-$, the $s$, $d_0$, and $d_{\pm 1}$ partial waves after the absorption of the first photon lead to final $p_0$ and $f_0$ waves for $^2P(m_t=0)$ target state and final $p_{\pm 1}$ and $f_{\pm 1}$ waves for $^2P(m_t=\mp 1)$. 
The RABBIT spectrum is again incoherent sum over the target states (we suppress the energy dependence):
\begin{equation}
P_2(\theta, \tau) = |A_{m_t=0}(\theta, \tau)|^2 + 2|A_{m_t=1}(\theta, \tau)|^2,
\end{equation}
where for the $^2P(m_t=0)$ target we have
\begin{equation}
\label{eq:clm_Amt0}
A_{m_t=0}(\theta, \tau) = A_1(\tau)\, Y_{10}(\theta) + A_3(\tau)\, Y_{30}(\theta)
\end{equation}
with
\begin{align}
\label{eq:clm_A1} A_1(\tau) &= a^+\, e^{i\omega\tau} + a^-\, e^{-i\omega\tau}, \\
\label{eq:clm_A1_P} a^{\pm} &= M^{\pm}_{s \to p_0} + M^{\pm}_{d_0 \to p_0},\\
A_3(\tau) &= M^+_{d_0 \to f_0}\, e^{i\omega\tau} + M^-_{d_0 \to f_0}\, e^{-i\omega\tau},
\end{align}
and for the $^2P(m_t=\mp 1)$ targets we get
\begin{equation}
\label{eq:clm_Amt1}
A_{m_t=-1}(\theta, \tau) = B_1(\tau)\, Y_{11}(\theta) + B_3(\tau)\, Y_{31}(\theta)
\end{equation}
with
\begin{align}
B_1(\tau) &= M^+_{d_1 \to p_1}\, e^{i\omega\tau} + M^-_{d_1 \to p_1}\, e^{-i\omega\tau}, \\
B_3(\tau) &= M^+_{d_1 \to f_1}\, e^{i\omega\tau} + M^-_{d_1 \to f_1}\, e^{-i\omega\tau}.
\end{align}
The key difference from the H$^-$ case is that with Cl$^-$ the final $p_0$ partial wave has two intermediate $s$ and $d_0$ contributions, as seen in Eqs.~\eqref{eq:clm_A1} and~\eqref{eq:clm_A1_P}. All other final  partial waves go only through the single $d$ intermediate wave.

As in the case of H$^-$, the Cl$^-$ cases which have a single intermediate $d$~wave lead to the RABBIT sideband oscillations for the corresponding final partial waves $\cos(2\omega\tau +\delta\varphi_d)$ with the same one-photon phase offset $\delta\varphi_d = \eta^-_d - \eta_d^+$, assuming sufficiently high electron energies where the continuum-continuum phases are constant and $\ell$-independent for the photodetachment.   
Here again, $\eta_d$ is the energy-dependent phase of the single-photon matrix elements from the initial $p$ orbital to final $d$~wave and $\eta_d^\pm=\eta_d(E\mp\omega)$.

For the final $p_0$~wave with $s$ and $d_0$ intermediate paths, the RABBIT phase offset is given by
$\delta\varphi_{p_0} = \arg(a^-) - \arg(a^+)$, where
\begin{equation}
\label{eq:clm_arg_sd}
\arg(a^{\pm}) = \eta_d^{\pm} + \arctan\left(\frac{|M_{s\to p_0}^{\pm}|\, \sin\Delta\eta^{\pm}}{|M_{d_0\to p_0}^{\pm}| + |M_{s\to p_0}^{\pm}|\, \cos\Delta\eta^{\pm}}\right)
\end{equation}
with $\Delta\eta^{\pm} = \eta_s(E_f \mp \omega) - \eta_d(E_f \mp \omega)$. Equation~\eqref{eq:clm_arg_sd} is similar to the one-photon phase for the $m_t=0$ channel, see Eq.~\eqref{eq:clm_1hv_delay}, but $\delta\varphi_{p_0}$ here is angle-independent and depends on the ratio of the two-photon matrix elements. The angular dependence of the RABBIT delay comes from weighting the cosine oscillations with $\delta\varphi_d$ and $\delta\varphi_{p_0}$ offsets by the angular factors in Eqs.~\eqref{eq:clm_Amt0} and~\eqref{eq:clm_Amt1}. 

In general, when multiple intermediate partial waves contribute, the angular dependence of the one-photon and RABBIT delays are not identical, even in the photodetachment case where the continuum-continuum phases do not contribute above 5--10~eV.  That fact is illustrated for Cl$^-$ in Fig.~\ref{fig:delays_chlorine_1d}a.

The difference in the angular shapes has two origins: (i) the angular geometric effect of different final partial waves and (ii) the magnitude effect where the $d/s$ ratio is in the two-photon case modified by the continuum-continuum transitions even at high energies. 
In the asymptotic region we can write the two-photon matrix elements as a product of single-photon and continuum-continuum terms, as in Appendices \ref{app:Born_RABBIT_amplitudes} and \ref{app:rabbitt_hydrogen_derivation}, 
\begin{equation}
\begin{aligned}
    M&^\pm_{\ell_i m_i \to \ell m \to \ell_f m_f}(E) \\ &\approx M^1_{\ell_i m_i\to \ell m}(E\mp\omega) C_{\ell m\to \ell_f m_f}(E\mp\omega, E) \, .
\end{aligned}
\label{eq:Mfactors}
\end{equation}
The ratio of the continuum-continuum terms $C_{\ell m\to \ell_f m_f}$ for different intermediate partial waves does not necessarily go to unity due to angular factors. For the Cl$^-$ case, $\vert C_{d_0\to p_0}\vert / \vert C_{s\to p_0}\vert \to 2/\sqrt{5}\approx 0.89$ as $E\to\infty$, which further modifies the ratio in the $\arctan$ function in Eq.~\eqref{eq:clm_arg_sd} in contrast to the one-photon case in Eq.~\eqref{eq:clm_1hv_delay}. 

At $\theta=0$ only the $p_0$ and $f_0$ final partial waves contribute to the RABBIT process [Eq.~\eqref{eq:clm_Amt0}]. The resulting RABBIT phase is numerically almost identical to the one-photon phase in Eq.~\eqref{eq:clm_1hv_delay}, where the $Y_{20}(\theta=0)/Y_{00}$ ratio is almost perfectly compensated by the continuum-continuum terms, $Y_{10}(\theta=0)$, and $Y_{30}(\theta=0)$, which leads to almost perfect agreement of the RABBIT and one-photon delays for $\theta=0$ as seen in Fig.~\ref{fig:delays_chlorine_1d}a. As the $\theta$~angle increases, the geometric and magnitude effects do not cancel each other any longer and the delays start to visibly differ.

In Fig.~\ref{fig:delays_chlorine_1d}b, we test the importance of the geometric and magnitude effects on the angular dependence. The RABBIT delay in Fig.~\ref{fig:delays_chlorine_1d}b (dashed) is calculated with two-photon matrix elements where we completely eliminated the effect of the continuum-continuum transitions, that is, we set 
$M^\pm_{\ell_i m_i \to \ell m \to \ell_f m_f}(E) = M^1_{\ell_i m_i\to \ell m}(E\mp\omega)$ but kept the correct two-photon angular dependence in Eqs.~\eqref{eq:clm_Amt0} and~\eqref{eq:clm_Amt1}. This RABBIT delay in Fig.~\ref{fig:delays_chlorine_1d}b is closer to the one-photon delay for a general angle than when the continuum-continuum transitions are taken into account as in Fig.~\ref{fig:delays_chlorine_1d}a. However, the disrepancy at $\theta=0$ is increased by omitting the magnitude effect. In the full calculation in Fig.~\ref{fig:delays_chlorine_1d}a, the largest differences are at $\sim$40~eV and $\sim$70~eV, where the $\vert M^1_{p\to d_0}\vert / \vert M^1_{p\to s}\vert$ ratio is close to one and the $\arctan$ in Eq.~\eqref{eq:clm_arg_sd} is most sensitive to the modification by the continuum-continuum ratio. A small difference of 2--5~as between the delays remains even at 100~eV at larger $\theta$ angles.

The RABBIT signal essentially vanishes close to $\theta=90^\circ$ (Fig.~\ref{fig:polar_plots_chlorine}b) resulting again in jumps by $\pi/(2\omega)$ in the RABBIT delay (Fig.~\ref{fig:polar_plots_chlorine}d) as discussed in Sec.~\ref{sec:hydrogen_dynamics}.

In the following section, we move on to analyze the photodetachment of NO$_3^-$, where we utilize the key results obtained from the H$^-$ and Cl$^-$ cases. Specifically, we have shown that when a single intermediate partial wave contributes (or dominates), the one-photon and RABBIT delays are isotropic and identical for photodetachment at electron energies above 5--10~eV.   We will see that this special case also applies to the nitrogen $1s$ detachment of NO$_3^-$ discussed in Sec.~\ref{sec:core_dynamics}.    On the other hand, the interference of two or more intermediate partial waves results in a nontrivial angular dependence, which is in general not identical for one-photon and RABBIT delays.  That is the behavior that applies for the cases of valence and oxygen $1s$ detachment of NO$_3^-$.

\section{\label{sec:one_photon_dynamics}Single-photon valence detachment dynamics of NO\boldmath{$_3^-$}}

\begin{figure}[t]
\includegraphics[scale=0.8]{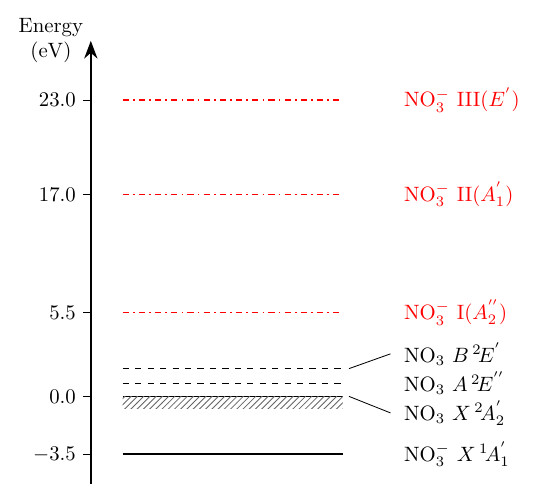}
\caption{\label{fig:energy_diagram} Energy diagram for valence photodetachment of NO$_3^-$ from its $X\,^1\!A_1^{'}$ ground state to the three lowest states of NO$_3$. The dot-dashed lines represent three shape resonances with different character of the temporarily-trapped electron. The energy axis is approximately in scale.}
\end{figure}

\begin{figure*}[]
\includegraphics[scale=1.0]{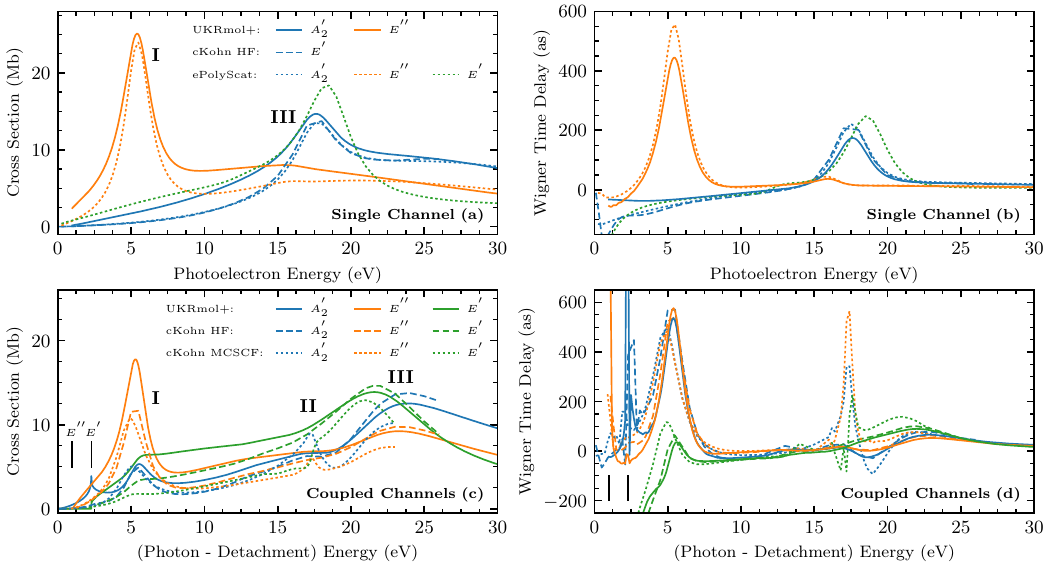}
\caption{\label{fig:total_xsec_delay} Comparison of calculated totally-averaged integral one-photon cross sections and corresponding Wigner time delays for the photodetachment of NO$_3^-$ to the three lowest states $X\,^2\!A_2^{'}$, $A\,^2\!E^{''}$, and $B\,^2\!E^{'}$ of NO$_3$ using our single- and coupled-channels models, which are summarized in Table~\ref{tab:models}. Single-channel results (panels a and~b) are plotted with respect to the photoelectron energy for each channel. Coupled-channel models (panels~c and~d) are shown for energies with respect to the $X\,^2\!A_2^{'}$ threshold. The position of the $A\,^2\!E^{''}$ and $B\,^2\!E^{'}$ thresholds in panels~c and~d are indicated by the vertical lines. Panel~b (d) shares the legend with panel~a (c). For the $A\,^2\!E^{''}$ and $B\,^2\!E^{'}$ channels, the cross sections are shown for only one degenerate component in panels~a and~b. To obtain the total cross sections, the results need to be multiplied by a factor of two. Resonances labeled I, II and III are described\ in the text.}
\end{figure*}

\begin{figure*}[]
\includegraphics[scale=1.0]{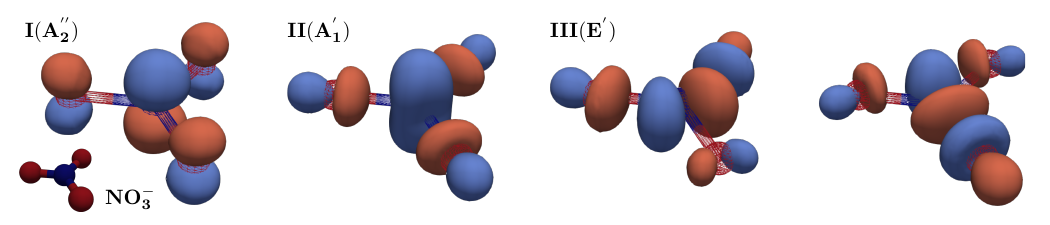}
\caption{\label{fig:resonant_orbitals} Orbitals for temporarily-attached electron representing the valence shape resonances I, II, and III of NO$_3^-$ with continuum symmetries $A_2^{''}$, $A_1^{'}$, and $E^{'}$, respectively. Both components of the doubly-degenerate $E^{'}$ orbital are shown. The geometry of the molecule is shown in the bottom left corner and additionally using wireframes in the orbital pictures.}
\end{figure*}

\begin{table}[h!]
\caption{\label{tab:models}Summary of single- and coupled-channel models for the one-photon valence detachment of NO$_3^-$.}
\begin{ruledtabular}
\begin{tabular}{lcp{0.3\textwidth}}
method & channels & orbitals \\
\hline
\textsc{UKRmol+} & 1 & HF orbitals for the $X\,^2\!A_2^{'}$ state of NO$_3$ with cc-pVDZ basis \\
\textsc{UKRmol+} & 3 & HF orbitals for the $X\,^2\!A_2^{'}$ state of NO$_3$ with cc-pVDZ basis \\
\textsc{ePolyScat} & 1 & HF orbitals for the $X\,^2\!A_2^{'}$ state of NO$_3$  with aug-cc-pVTZ basis \\
\textsc{cKohn} & 1 & HF orbitals for the $X\,^2\!A_2^{'}$ state of NO$_3$ with Dunning-Hay + diffuse basis\\
\textsc{cKohn} & 3 &  MCSCF orbitals averaged over  the $X\,^2\!A_2^{'}$, $A\,^2\!E''$, and $B\,^2\!E'$ states of NO$_3$ with Dunning-Hay + diffuse basis\\
\end{tabular}
\end{ruledtabular}
\end{table}

\begin{table}[]
\caption{\label{tab:resonances} Symmetry considerations for the appearance of shape resonances in the valence and core photodetachment of NO$_3^-$. The table lists the character of the detaching electron, together with the symmetries of the dipole operator, target state left behind, and the continuum symmetries containing the resonances. The resonances are labeled by Roman numerals according to their energy (I is the lowest resonance), independently for the valence, nitrogen~$1s$, and oxygen~$1s$ cases. The last column gives the dominant partial waves $\ell$ of the resonances ($\ell$s shown in the parentheses are not dominant but still significantly contribute). The initial state of NO$_3^-$ is always the ground state, which is totally symmetric.}
\begin{ruledtabular}
\begin{tabular}{lllllll}
character & dipole & target & cont. & res. & $\ell$ \\
\hline
valence & $E^{'}(x,y)$ & $^2\!A_2^{'}$ & $^2\!E^{'}$ & III & 4,(1,2) \\
valence & $E^{'}(x,y)$ & $^2\!E^{''}$ & $^2\!A_2^{''}$ & I & 1,(3) \\ 
valence & $A_2^{''}(z)$ & $^2\!E^{''}$ & $^2\!E^{'}$ & III & 3,(1) \\
valence & $E^{'}(x,y)$ & $^2\!E^{'}$ & $^2\!A_1^{'}$ & II & --- \\
valence & $E^{'}(x,y)$ & $^2\!E^{'}$ & $^2\!E^{'}$ & III & --- \\
nitrogen 1s & $E^{'}(x,y)$ & $^2\!A_1^{'}$ & $^2\!E^{'}$ & I & 2\\
oxygen 1s & $E^{'}(x,y)$ & $^2\!A_1^{'}$ & $^2\!E^{'}$ & I & 1\\
oxygen 1s & $E^{'}(x,y)$ & $^2\!A_1^{'}$ & $^2\!E^{'}$ & II & 1,2,3\\
oxygen 1s & $E^{'}(x,y)$ & $^2\!E^{'}$ & $^2\!E^{'}$ & I & 0,1,2\\
oxygen 1s & $E^{'}(x,y)$ & $^2\!E^{'}$ & $^2\!E^{'}$ & II & 1,2,3,4\\
\end{tabular}
\end{ruledtabular}
\end{table}

\begin{table*}[]
\caption{\label{tab:detachment_energies}Electron detachment energies for the three lowest states of NO$_3$ from the $X\,^1\!A_1^{'}$ ground state of NO$_3^-$ calculated with our models. The values are in electronvolts.}
\begin{ruledtabular}
\begin{tabular}{llllllll}
target & \textsc{UKRmol+} (1-chn) & \textsc{UKRmol+} (3-chn) & \textsc{ePolyScat} (1-chn) & \textsc{cKohn} (1-chn)\footnote{\label{ip_table_footnote}With MCSCF orbitals.} &\textsc{cKohn} (3-chn)\footnotemark[1] & Exp.\footnote{Experimental adiabatic electron detachment energies from Ref.~\cite{wange2002}.}\\
\hline
$X\,^2\!A_2^{'}$ & 3.78 & 5.31 & 3.50\footnote{\label{ip_table_footnote2}Manually set to 3.5~eV.} & & 3.50\footnotemark[3] & 3.92\\
$A\,^2\!E^{''}$ & 5.30  & 6.29 & 4.43 & & 4.51 & 4.78\\ 
$B\,^2\!E^{'}$ & --- & 7.60 & 5.52 & & 5.83 & 5.78\\
\end{tabular}
\end{ruledtabular}
\end{table*}

In this section, we turn to our calculations for the valence detachment dynamics of NO$_3^-$ triggered by a single photon. We find that valence detachment from NO$_3^-$ has two pronounced shape resonances at 5.5~eV and 23~eV above the first detachment threshold.

The nitrate anion, NO$_3^-$, is a molecular anion with a $X\,^1\!A_1^{'}$ closed-shell ground electronic state with the trigonal planar symmetry $D_{3h}$. In all our calculations, NO bonds of 1.22~\AA{} and ONO angles of 120$^\circ$ from a geometry optimization at the Hartree-Fock level were considered  -- very close to the experimental bond length of 1.2377~\AA{} for the equilibrium geometry of the anion~\cite{kawaguchi1998}. 
The neutral NO$_3$ has an open-shell $X\,^2\!A_2^{'}$ ground electronic state and two doubly-degenerate low-lying excited states $A\,^2\!E^{''}$ and $B\,^2\!E^{'}$, as sketched in the energy diagram in Fig.~\ref{fig:energy_diagram}~\cite{weaver_examination_1991, wange2002}.

The primary purpose of this work is to study the effect of the measurement (the interaction with the IR field) on the time delays using the R-matrix time-independent and time-dependent methods~\cite{masin2020_ukrmolp, houfek2024, benda2021, benda2022, brown2020_rmt}. Since solving the TDSE is computationally costly, we consider simple single- and coupled-channel models with Hartree-Fock (HF) orbitals for the bound electrons in the R-matrix calculations. To verify that these \textsc{UKRmol+} models give qualitatively good results for the single-photon dynamics, we also performed both single-channel Schwinger~\cite{Gianturco1994, Natalense1999} (\textsc{ePolyScat}~\cite{Schneider2020}) and single-channel complex-Kohn calculations (cKohn HF~\cite{KohnPhotoionization_1993, rlm95, Miyabe_CO2_2009, Rescigno_Douget_Orel_2012, trevisan_2012}) calculations with higher-quality HF orbitals. Coupled-channel complex Kohn calculations  using either uncorrelated HF orbitals (cKohn HF) or  correlated MCSCF orbitals (cKohn MCSCF) were also carried out to compare with the R-matrix calculations. The single-channel calculations considered each target channel ($X\,^2\!A_2^{'}$, $A\,^2\!E^{''}$, and $B\,^2\!E^{'}$) separately,\footnote{All R-matrix and complex Kohn calculations were performed using the abelian $C_{2v}$ subgroup of $D_{3h}$. In $C_{2v}$, $E^{'}$ and $E^{''}$ representations of $D_{3h}$ are no longer irreducible and transform as $A_1+B_2$ and $A_2+B_1$ representations of $C_{2v}$, respectively. All calculations in $C_{2v}$ always included both degenerate components of the $E$ states to preserve the symmetry. The Schwinger calculations using \textsc{ePolyScat} were performed in the $D_{3h}$ symmetry.} while the coupled-channel models included
these three channels at the same time in the close-coupling expansion of the total final scattering function given by Eq.~\eqref{eq:close_coupling}.

In all R-matrix calculations, we used HF orbitals calculated for the $X\,^2\!A_2^{'}$ state of NO$_3$ with the cc-pVDZ basis set~\cite{dunning1989} using \textsc{MOLPRO}~\cite{molpro_2012, molpro_2020, molpro_version_2022d3}. The target wave functions $\Phi_\alpha$ were described by single HF configurations in both single- and coupled-channel models. The basis set for the continuum electron consisted of 160 ninth-order B-splines with maximum angular momentum $\ell_\mathrm{max}=6$ in the R-matrix sphere with radius~$R$ of 80~bohr. To ensure that the cross sections, one-photon and RABBIT delays are converged with respect to the R-matrix radius and $\ell_\mathrm{max}$, we performed a limited number of calculations using $R=40,80,120$~bohr with $\ell_\mathrm{max}=6$ and $R=40$~bohr with $\ell_\mathrm{max}=9$, resulting in maximum error of RABBIT delays, which are the most sensitive to these numerical parameters, around 3\% at the 5.5-eV resonance and up to 5~as elsewhere. 

The scattering calculations using \textsc{ePolyScat} were also based on restricted Hartree-Fock (RHF) orbitals, obtained using \textsc{MOLPRO} \cite{molpro_version_2022d3}, with, however,  an aug-ccVTZ basis set~\cite{dunning1989,Kendall1992}.  The photodetachment cross sections and Wigner time delays were computed using ePolyScat \cite{Gianturco1994,Natalense1999} in the full point-group symmetry of the molecule, using just a single scattering channel with a single-configuration representation for the target states. The single-center expansions in \textsc{ePolyScat} used up to $\ell_\mathrm{max}=80$.

For the complex Kohn calculations, we used a Dunning-Hay TZP basis, augmented with two $s$, three $p$ and two $d$ diffuse functions. For the single-channel calculations, the cKohn and \textsc{ePolyScat} give essentially identical results, shown explicitly for the $E^{'}$ state in Fig.~\ref{fig:total_xsec_delay}a, attesting to the adequacy of the basis sets employed in the cKohn and \textsc{ePolyScat} calculations.

In the single-channel models with HF orbitals, the target states were represented using single configurations. In this static-exchange approximation, no polarizability is taken into account so that the molecular potential exponentially decreases with the distance. Even though the same single-configuration representations were used in the coupled-channel models with HF orbitals, the latter calculations include some polarization due to the coupling between the three channels with optically allowed transitions.\footnote{\label{footnote:target_dipole_coupl}Dipole transitions between $E^{''}$ and $E^{'}$ states are symmetry allowed, as well as transitions between $A_2^{'}$ and $E^{'}$.} The \textsc{cKohn} coupled-channel model with the MCSCF orbitals describes the polarization effects the best among our models due to both channel coupling and included correlating configurations~\cite{KohnPhotoionization_1993, rlm95, Miyabe_CO2_2009, Rescigno_Douget_Orel_2012, trevisan_2012}. 

The employed time-independent methods and the manner in which they were used are summarized in Table~\ref{tab:models}. The photodetachment cross sections and Wigner time delay averaged over molecular orientations and integrated over electron emission angles are shown in Fig.~\ref{fig:total_xsec_delay}.

Beginning with the single-channel cases, we observe two pronounced shape resonances, labeled as~I and~III, in the cross sections (Fig.~\ref{fig:total_xsec_delay}a).
Resonance~I, where the temporarily-attached electron has the $A_2^{''}$ symmetry, appears at electron energy of 5.5~eV in the $A\,^2\!E^{''}$ channel. It leads to the averaged Wigner time delay of 500~as. Resonance~III with the continuum electron of the $E^{'}$ symmetry appears in the 
$X\,^2\!A_2^{'}$ and $B\,^2\!E^{'}$ channels at approximately 18~eV for the electron energy with Wigner delay around 200~as (Figs.~\ref{fig:total_xsec_delay}a and~\ref{fig:total_xsec_delay}b). Resonance~III also weakly appears in the $A\,^2\!E^{''}$ channel at somewhat lower energy of 16 eV, see the small peak at 16~eV in Fig.~\ref{fig:total_xsec_delay}b. The much smaller Gaussian basis cc-pVDZ used for the bound electrons in the R-matrix calculations is the reason for the quantitative difference of \textsc{UKRmol+} single-channel calculation in comparison to the \text{ePolyScat} and \textsc{cKohn} results in Figs.~\ref{fig:total_xsec_delay}a and~\ref{fig:total_xsec_delay}b. However, the \textsc{UKRmol+} model qualitatively reproduces the \textsc{ePolyScat} and \textsc{cKohn} results very well, suggesting that this simpler model is sufficient for the semi-quantitative calculation of RABBIT delays.

Figure~\ref{fig:resonant_orbitals} shows orbitals for the electron temporarily trapped behind the centrifugal barriers forming the shape resonances I, II, and III of the three different characters. The resonant positions are also illustrated in the energy diagram in Fig.~\ref{fig:energy_diagram}. Note that although the resonant one-electron orbital is nearly the same for all target-state channels, adding this orbital to different target states leads to  different $(N+1)$-electronic resonant states. 
Table~\ref{tab:resonances} summarizes the appearance of the shape resonances in the target-state channels.
The orbitals shown in Figure~\ref{fig:resonant_orbitals} are Siegert states~\cite{Korsch1984} computed using a local potential, truncated at $r=9.5$~\AA, which is an approximation to the static-exchange potential \cite{Natalense1999} that was used in the \textsc{ePolyScat} ionization calculations. The resonance I($A_2^{''}$) is a $\pi^*$ resonance at $\sim$5 eV, while resonance II($A_1^{'}$) is a $\sigma^*$ resonance which is very weak in the photodetachement cross section, and finally resonance III(E$^{'}$) is a stronger $\sigma^*$ resonance with higher angular momenta as indicated in Table~\ref{tab:resonances}.

Turning to the coupled-channel calculations, with results summarized in Figs.~\ref{fig:total_xsec_delay}c and~\ref{fig:total_xsec_delay}d, we again find resonance I centered near $\sim$5 eV and resonance III, shifted several eV upward relative to the single-calculations. For a better comparison, the data are shown as a function of energy that is measured with respect to the $X\,^2\!A_2^{'}$ threshold to align the channel features in terms of photon energy. Table~\ref{tab:detachment_energies} lists calculated detachment energies in comparison to the experimental values. We now find a third feature, which we label resonance II, near 17 eV. However, this feature only figures prominently in the coupled-channel \textsc{cKohn} calculations and is completely missing in the single-channel calculations.

Due to the interchannel coupling, resonance~I, originally only in the $A\,^2\!E^{''}$ channel without channel coupling, appears in all three channels and is shifted slightly lower in energy. 
The $X\,^2\!A_2^{'}$ channel strongly couples to $A\,^2\!E^{''}$ resulting in basically the same shape and width of the resonance peak in the  $X\,^2\!A_2^{'}$ channel, which consequently leads to almost the same Wigner delay of 550~as, visible in Figs.~\ref{fig:total_xsec_delay}c and~\ref{fig:total_xsec_delay}d. The $B\,^2\!E^{'}$ channel is coupled to $A\,^2\!E^{''}$ much more weakly resulting in a shoulder in the cross section instead of a prominent peak (Fig.~\ref{fig:total_xsec_delay}c). Furthermore, the delay for the $B\,^2\!E^{'}$ channel is much shorter, only around 20--50~as, due to the interaction with the background (non-resonant) contribution to the photodetachment process. Resonance~III is broadened and shifted to slightly higher energies by the interchannel effects and is visible in the  $A\,^2\!E^{''}$ channel as well (Fig.~\ref{fig:total_xsec_delay}c). The broadening results in a shorter Wigner delay around 100~as, approximately a half of the single-channel results (Fig.~\ref{fig:total_xsec_delay}d). Additionally, we observe a dramatic behavior of the Wigner delay in the $X\,^2\!A_2^{'}$ channel at the opening of the $B\,^2\!E^{'}$ channel. As for the resonance II feature found in the \textsc{cKohn} calculations, 
detailed  analysis of the results showed that this minor peak comes from excitations out of inner-valence orbitals, showing that resonance II arises, not from channel coupling, but from employing correlated wave functions in the calculations. However, since no inner-valence states were explicitly treated in any of the calculations, it is unclear whether these features are physical or rather pseudostates.

As in the case of the single-channel calculations, the coupled-channel \textsc{UKRmol+} model with the small Gaussian basis gives qualitatively similar results as the \textsc{cKohn} model with higher-quality neutral HF orbitals.

In this section, we have described the main features of the electron dynamics when a valence electron is detached by a single photon. We now turn our attention to the measurability of these single-photon time delays, that is, we consider the interaction of the system with an additional IR photon to model RABBIT and streaking processes.

\section{\label{sec:two_photon_dynamics_LF}Two-photon valence detachment dynamics of NO\boldmath{$_3^-$} in Lab Frame}

\begin{figure*}[]
\includegraphics[width=1.0\textwidth]{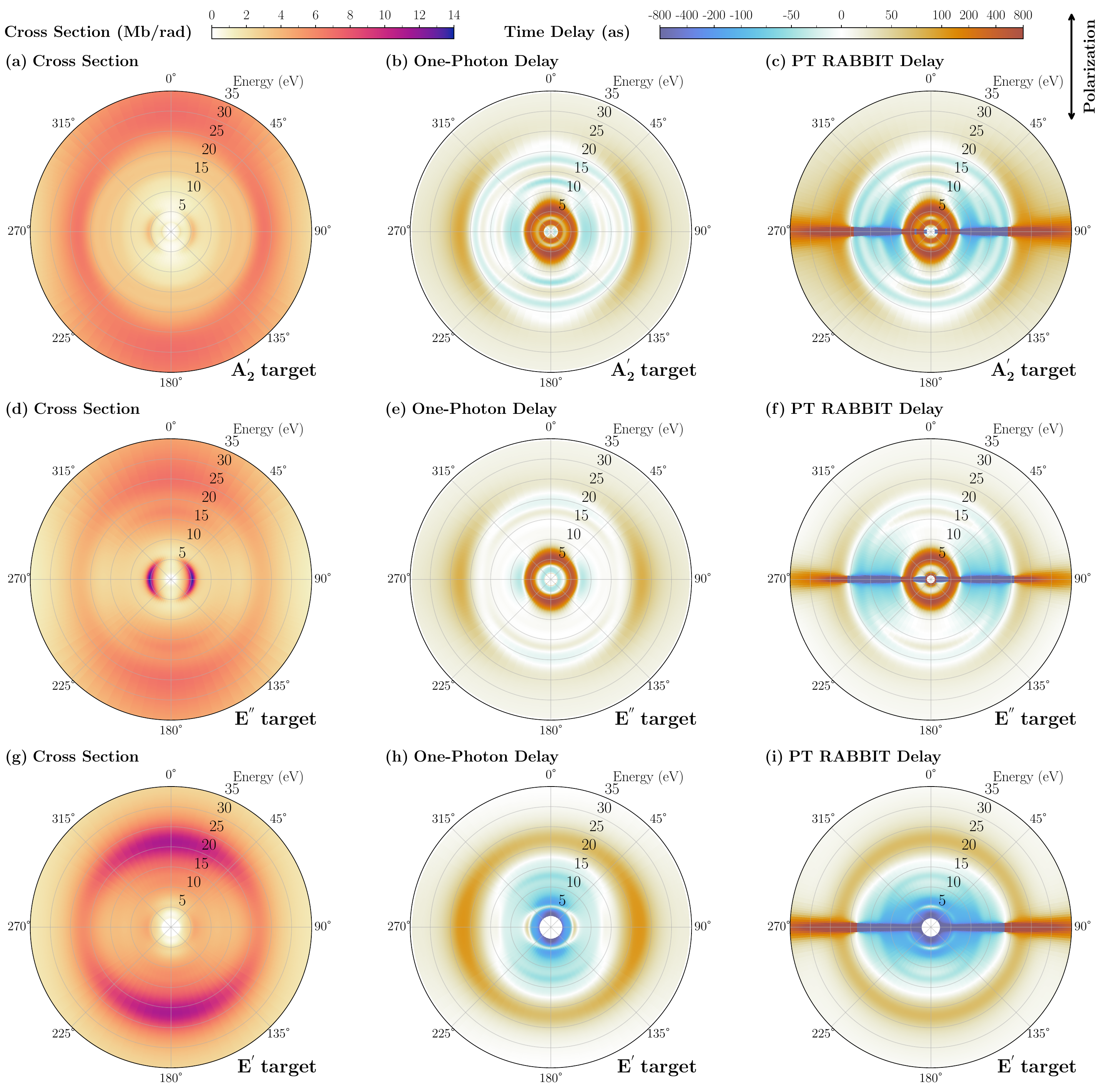}
\caption{\label{fig:polar_plots_lab_frame}
One-photon cross section (first column), one-photon time delays (second column), and perturbation-theory (PT) RABBIT time delays (third column) for the valence photodetachment of NO$_3^-$ plotted as a function of electron energy (radial direction) and angle between the polarization direction (vertical) and electron emission polar angle in the laboratory frame. The first, second, and third rows show the results for the $X\,^2\!A_2^{'}$, $A\,^2\!E^{''}$, and $B\,^2\!E^{'}$ final states of NO$_3$, respectively. The data were calculated using the \textsc{UKRmol+} coupled-channel model. The IR energy is 0.517~eV (2.4~$\mu$m). The time delays are plotted using a symmetric logarithmic scale with a linear scale from -100 to 100~as. The electron energy in all panels is measured with respect to the $X\,^2\!A_2^{'}$ ground state of NO$_3$ (the same as in Figs.~\ref{fig:total_xsec_delay}c and~\ref{fig:total_xsec_delay}d).
}
\end{figure*}

\begin{figure}[]
\includegraphics[scale=1.0]{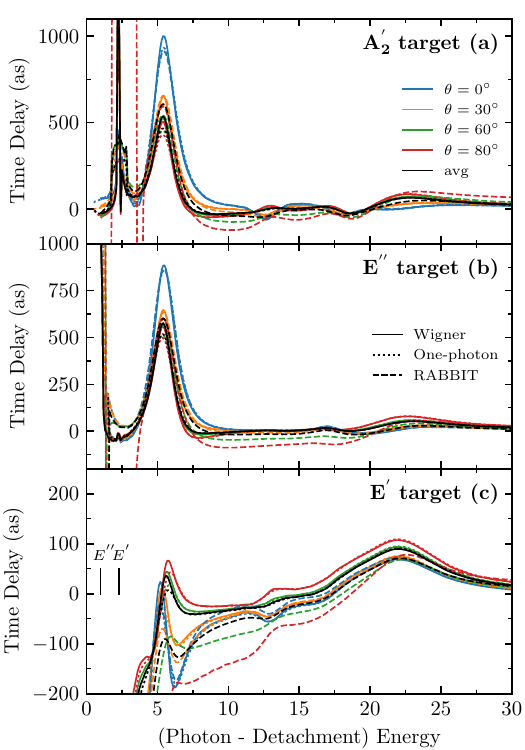}
\caption{\label{fig:1hv_2hv_1d_lab_frame}
Comparison of Wigner (solid lines), one-photon (dotted), and perturbation-theory RABBIT (dashed) time delays for valence detachment of
NO$_3^-$ in the laboratory frame. The results calculated with the \textsc{UKRmol+} coupled-channel model are shown for electron emission polar angles of 0, 30, 60, and 80 degrees and also averaged over the polar angle (labeled as \emph{avg}) 
for the $X\,^2\!A_2^{'}$, $A\,^2\!E^{''}$, and $B\,^2\!E^{'}$ states of NO$_3$ in panels a, b, and~c, respectively. 
The IR energy is 0.517~eV (2.4~$\mu$m). The legend is shared among all panels and the energy is measured with respect to the $X\,^2\!A_2^{'}$ threshold.}
\end{figure}

We first focus on the two-photon valence dynamics of NO$_3^-$ in the laboratory frame, that is with fixed angle between the light polarization direction and the electron emission direction, averaged over all molecular orientations.
In particular, we here compare the one-photon time delays with perturbation-theory RABBIT delays.

In all our RABBIT calculations presented here, we used the same \textsc{UKRmol+} coupled-channel model of the molecular system as in the single-photon case described in Sec.~\ref{sec:one_photon_dynamics} with the R-matrix radius of $R=80$~bohr and $\ell_\mathrm{max}=6$. 

In our RABBIT calculations here and in the other sections, we consider an IR energy of 0.517~eV corresponding to a wavelength of 2.4~$\mu$m for two reasons: (i) to avoid IR-induced transitions~\cite{benda2022, benda_dipole-laser_2024, delgado_three-path_2025} among the target states of neutral NO$_3$, which are separated by about 1~eV, and (ii) to achieve a better energy resolution in the time-dependent RABBIT simulations presented in Sec.~\ref{sec:two_photon_dynamics_MF}.
Although most of the time delay measurements have been done using 800~nm IR wavelegths from Ti:sapphire lasers, 
longer IR wavelengths in the 1-2~$\mu$m range have been used as well~\cite{schoun_attosecond_2014, weber_flexible_2015, saito_attosecond_2016,cousin_attosecond_2017,heinrich_attosecond_2021}. We consider the same polarization for the high-frequency and IR fields in all RABBIT calculations.
Recall that the ``one-photon'' delay from RABBIT calculations (or measurements) here means the finite-difference approximation to the Wigner delay using the harmonic energies,  as in  Eq.~\eqref{eq:1hv_delay}.

Figure~\ref{fig:polar_plots_lab_frame} presents the single-photon photodetachment cross sections, together with one-photon and RABBIT time delays for the valence detachment of NO$_3^-$ as a function of energy and electron polar angle in the laboratory frame for all three low-lying states of NO$_3$.

For the $X\,^2\!A_2^{'}$ and $A\,^2\!E^{''}$ channels, resonance~I at 5.5~eV dominates the dynamics and the one-photon time delay
is almost isotropic, reaching 900--1000~as along the polarization direction ($\theta=0^\circ,180^\circ$) and gradually decreasing to 500--600~as for angles $30^\circ < \theta < 150^\circ$ (Figs.~\ref{fig:polar_plots_lab_frame}b and~\ref{fig:polar_plots_lab_frame}e).
One-dimensional delay lineouts for selected angles are also shown in Fig.~\ref{fig:1hv_2hv_1d_lab_frame}. Note that the one-photon delay is essentially the same as the Wigner delay in Fig.~\ref{fig:1hv_2hv_1d_lab_frame} for all energies except at narrow structures. 
Resonance~I is weakly coupled to the $B\,^2\!E^{'}$ channel and the one-photon delay is not isotropic due to the interaction with the background, reaching -20~as along the polarization direction and 40~as at 90$^\circ$ (Figs.~\ref{fig:polar_plots_lab_frame}g and~\ref{fig:1hv_2hv_1d_lab_frame}c). In the case of resonance~III at $\sim$23~eV, the situation is opposite. Resonance~III predominantly appears in the $B\,^2\!E^{'}$ channel with almost isotropic delay (Figs.~\ref{fig:polar_plots_lab_frame}h and~Fig.~\ref{fig:1hv_2hv_1d_lab_frame}c) of 75~as for $\theta=0^\circ,180^\circ$ and 110~as for $\theta=90^\circ$. In the $X\,^2\!A_2^{'}$ and $A\,^2\!E^{''}$ channels, resonance~III leads to almost zero delay along the polarization axis and to 80~as at $\theta=90^\circ$ (Figs.~\ref{fig:polar_plots_lab_frame}b, \ref{fig:polar_plots_lab_frame}e, \ref{fig:1hv_2hv_1d_lab_frame}a, and~\ref{fig:1hv_2hv_1d_lab_frame}b). The energy region between resonances~I and~III (6--20~eV) show some structure with short one-photon delays varying from -60~as to 30~as originating from the energy dependence of the dominant partial waves and also interference of other contributing partial waves at these energies. 

The perturbation-theory RABBIT time delays for all three channels are shown in the third column of Fig.~\ref{fig:polar_plots_lab_frame} and as dashed lines in Fig.~\ref{fig:1hv_2hv_1d_lab_frame}. In general, the RABBIT delay closely follows the one-photon delay for energies above 4~eV in our calculations, and qualitatively, 
the angular dependence behaves the same as in the chlorine anion discussed in Sec.~\ref{sec:chlorine_dynamics}. At electron emission angles close to the IR polarization axis ($\theta=0^\circ,180^\circ$), the one-photon and RABBIT delays are almost identical but as the electron leaves further away from the polarization axis, the differences between the delays increases (Fig.~\ref{fig:1hv_2hv_1d_lab_frame}). This is most notable for the $B\,^2\!E^{'}$ channel, where at 10~eV the delay difference is around 80~as at $\theta=60^\circ$ and almost 150~as at $\theta=80^\circ$.
Near $\theta=90^\circ$, the RABBIT signal shown in Figs.~1--3 in the SM~\cite{SM} again effectively vanishes and the $\pi/(2\omega)$ jumps appear in the RABBIT delay.  See Sec.~\ref{sec:hydrogen_dynamics} and Ref.~\cite{busto_fanos_2019} for  discussions of this behavior. 
The SM~\cite{SM} includes polar plots (Figs.~4--6) showing the difference between the RABBIT and one-photon delays.
As a result of the angular dependence, the delays averaged over the emission angle (black curves in Fig.~\ref{fig:1hv_2hv_1d_lab_frame}) show a difference of 20--25~as even at the position of resonance~III at 23~eV. Note that the dramatic behavior of the RABBIT delays near $\theta=90^\circ$ does not affect the averaging because the RABBIT signal is averaged, not the delays, as discussed in Appendix~\ref{app:lab_frame}. 

As in the Cl$^-$ case described in Sec.~\ref{sec:chlorine_dynamics}, the interference of several intermediate partial waves gives rise to differences  between angular dependences of the RABBIT and Wigner delays. In the molecular case, the initial state already contains many different partial waves due to the non-sphericity of the molecule. The dipole transitions lead to several intermediate partial waves significantly contributing to the RABBIT dynamics. 

The effect of the molecular potential on the continuum-continuum transitions is limited to very low energies below 4--5~eV, which is in a good agreement with the behavior seen in calculations for atomic F$^-$ and Cl$^-$ by Lindroth and Dahlstr\"{o}m~\cite{lindroth_attosecond_2017}. Note that in contrast to Ref.~\cite{lindroth_attosecond_2017}, our Cl$^-$ calculation does not include polarization effects. Our coupled-channel calculations for NO$_3^-$ partially include polarization effects  as mentioned in Sec.~\ref{sec:one_photon_dynamics}, and despite it, the continuum-continuum delay is negligible at the 5.5-eV resonance. More specifically, the difference between RABBIT and one-photon delay at 5.5~eV and $\theta=0^\circ$ is only -20, 5, and 10~as for the $X\,^2\!A_2^{'}$, $A\,^2\!E^{''}$,  $B\,^2\!E^{'}$ channels, respectively.

Hence we see that while there are large differences in the angular dependence of the RABBIT delay and the Wigner delay, a RABBIT calculation along the polarization direction of the high-frequency and IR fields is very close to the underlying Wigner delay.  The long delays at resonances as well as nonresonant delays would be directly accessible to such a measurement.

\section{\label{sec:two_photon_dynamics_MF}Two-photon valence detachment dynamics of NO\boldmath{$_3^-$} in Molecular Frame}

\begin{figure*}[]
\includegraphics[width=1.0\textwidth]{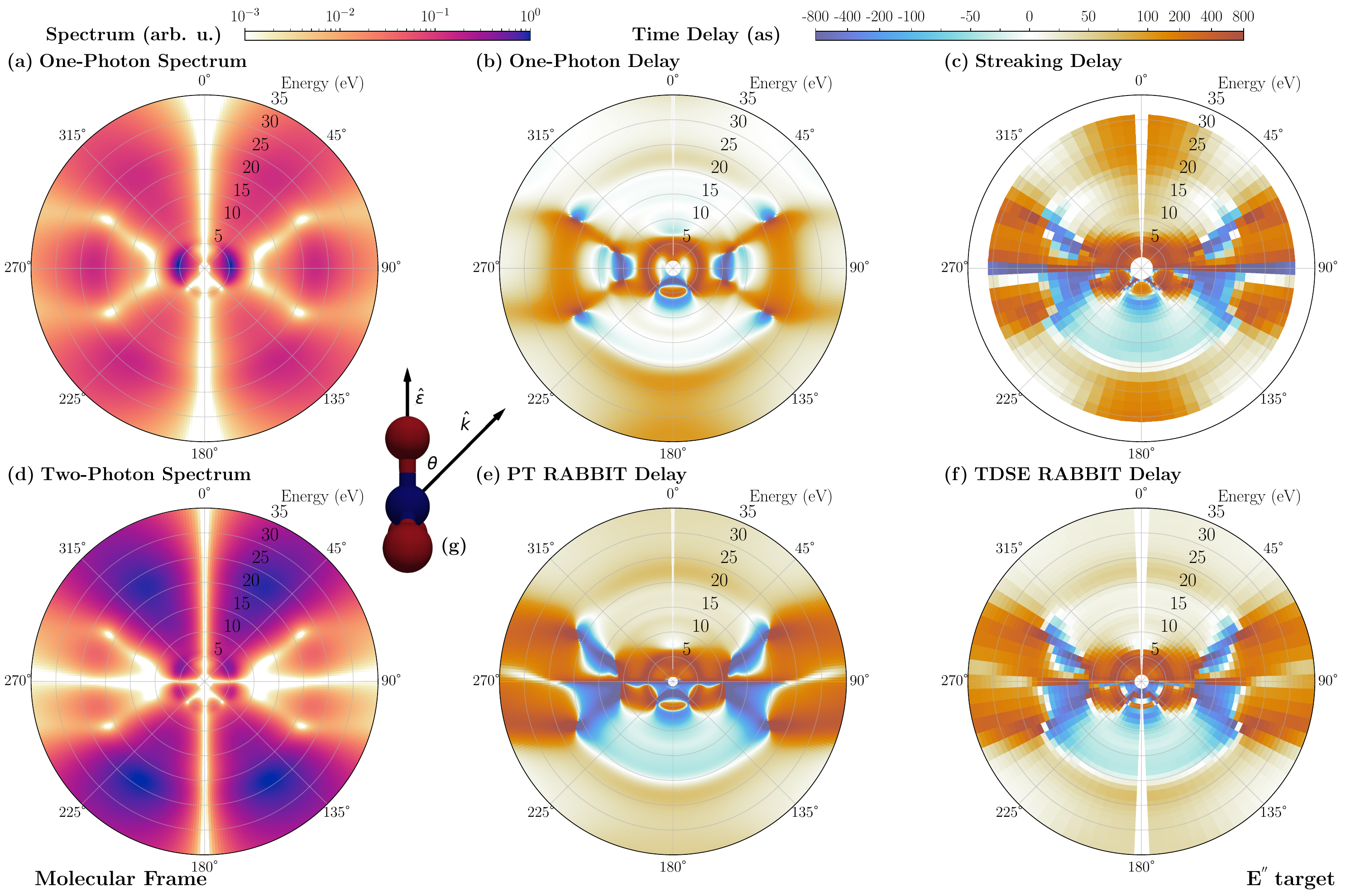}
\caption{\label{fig:delay_mf_Epp}
Valence photodetachment of NO$_3^-$ in the molecular frame with the neutral molecule left in the $A\,^2\!E^{''}$ state.
Panels a and~d show the one- and two-photon spectra in a logarithmic scale, panels~b and~e show the one-photon and perturbation-theory (PT) RABBIT delays, and finally, panels~c and~f show the results of time-dependent simulations of the streaking and RABBIT processes, respectively. 
The data calculated with the \textsc{UKRmol+} coupled-channel model are shown as a function of the electron energy measured with respect to the $X\,^2\!A_2^{'}$ state and electron emission polar angle~$\theta$. 
Both, the high-frequency and IR fields have a linear polarization~$\hat{\varepsilon}$ along one of the NO bonds and the electron is emitted in the plane perpendicular to the molecular plane, see inset~g. The IR energy is 0.517~eV (2.4~$\mu$m). The delays are plotted using a symmetric logarithmic scale with linear scale between -100 and 100~as.
}
\end{figure*}

\begin{figure}[]
\includegraphics[width=0.5\textwidth]{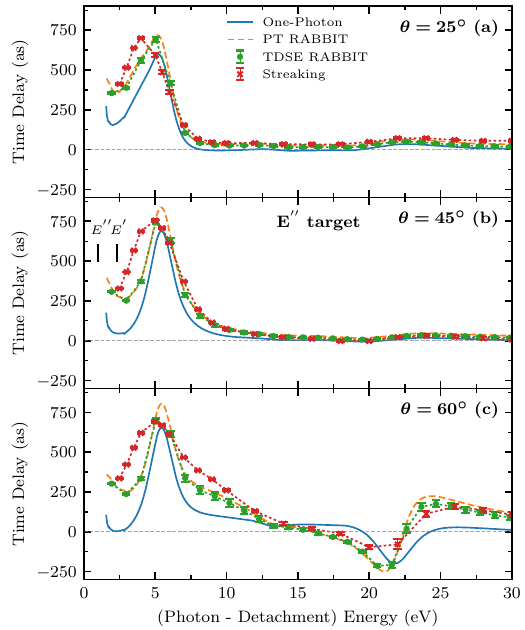}
\caption{\label{fig:delay_mf_Epp_1d}
Comparison of one-photon, perturbation-theory (PT) and TDSE RABBIT, and streaking time delays for valence detachment of NO$_3^-$ into
the final state $A\,^2\!E^{''}$ of NO$_3$ in the molecular frame for electron polar angles of 25, 45, and 60 degrees in panels a, b, and~c, respectively. The electron and polarization directions are illustrated in Fig.~\ref{fig:delay_mf_Epp}g.
The data were calculated using the \textsc{UKRmol+} coupled-channel model and the energy is measured with respect to the ground $X\,^2\!A_2^{'}$ state. The error bars are given by fitting errors.
} 
\end{figure}

\begin{figure}[]
\includegraphics[width=0.5\textwidth]{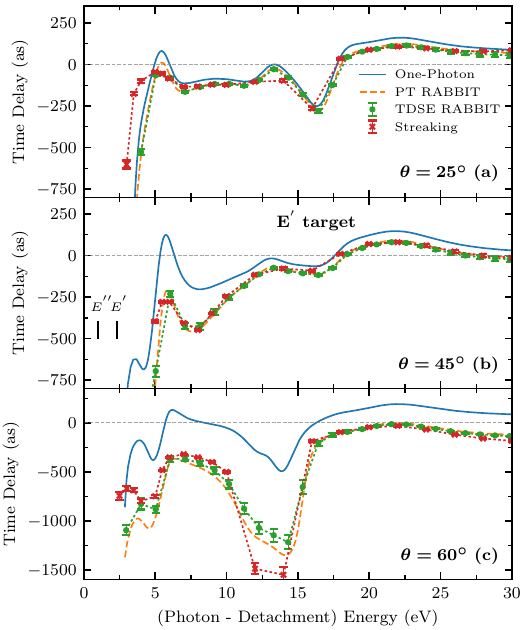}
\caption{\label{fig:delay_mf_Ep_1d}
Comparison of one-photon, perturbation-theory (PT) and TDSE RABBIT, and streaking time delays for valence detachment of NO$_3^-$ into
the final state $B\,^2\!E^{'}$ of NO$_3$ in the molecular frame for electron polar angles of 25, 45, and 60 degrees in panels a, b, and~c, respectively. The electron and polarization directions are illustrated in Fig.~\ref{fig:delay_mf_Epp}g.
The data were calculated using the \textsc{UKRmol+} coupled-channel model and the energy is measured with respect to the ground $X\,^2\!A_2^{'}$ state. The error bars are given by fitting errors.
} 
\end{figure}

In this section, we discuss results for valence photodetachment dynamics of NO$_3^-$ in the molecular frame. Beyond the one- and two-photon time-independent calculations, we also performed time-dependent simulations of the RABBIT and streaking processes using the R-matrix theory suite of codes to solve the TDSE. 

For the time-dependent RABBIT process, we modeled the electric field for the attosecond pulse train (APT) of odd harmonics using formulas of Serov and Kheifets~\cite{serov_time_2017} with IR energy of 0.517~eV (2.4~$\mu$m) and 13 IR cycles. The period of one IR cycle is 8~fs. The spectrum (Fourier transform) of the APT electric field is shown in Fig.~7 in the SM~\cite{SM}. The resulting APT field with peak intensity around 10$^{10}$~W/cm$^2$ was combined with the IR field with peak intensity of 10$^9$~W/cm$^2$. These two fields were mutually delayed at 19  different delays covering more than a half of one IR cycle (around 6~fs)
to well resolve one sideband oscillation with a 4~fs period. The oscillations were fitted by the cosine formula in Eq.~\eqref{eq:rabitt_signal_cos} to determine the RABBIT time delay for each sideband energy and electron emission angle. The system was propagated for the duration of the pulses (about 100~fs) and a large box of 8000~bohr was used to fully contain the electron wave packets. 

In the case of the streaking simulations, a single pulse with duration of 1~fs (full width half maximum of the field intensity), intensity of 10$^{11}$~W/cm$^2$, and central energies ranging from 8 to 35~eV was combined with one-cycle IR with energy of 0.517~eV and 10$^9$~W/cm$^2$ at 27 mutual time delays to resolve the streaking traces, which were fitted by Eq.~\eqref{eq:streaking} to extract the streaking time delay.

The time-dependent calculations used the same \textsc{UKRmol+} coupled-channel model for the description of the field-free molecular system.
For the molecular-frame calculations, we considered both field polarizations to be along one NO bond and electrons emitted in the plane that is perpendicular to the molecular plane and contains the mentioned NO bond, see inset~g in Fig.~\ref{fig:delay_mf_Epp}. In this configuration, the spectrum for the $X\,^2\!A_2^{'}$ state has a nodal plane. The resulting one- and two-photon spectra and four kinds of delays (one-photon, perturbation-theory and TDSE RABBIT, and streaking) are shown for the $A\,^2\!E^{''}$ target state in Fig.~\ref{fig:delay_mf_Epp} and for the $B\,^2\!E^{'}$ state in Fig.~8 in the SM~\cite{SM}.

The spectra and time delays in Fig.~\ref{fig:delay_mf_Epp} are no longer symmetrical along the polarization axis ($\theta=0^\circ$ and 180$^\circ$) due to the molecular asymmetry. The delays show more complex structure than in the laboratory frame presented in the previous section. The interference of the electron partial waves, which is not smeared out by the averaging over molecular orientation, leads to the signal vanishing for certain directions and energies, which causes rapid changes in the time delays as the trajectory of the dipole matrix elements in the complex plane as a function of energy approaches the origin. The RABBIT delays from the time-independent and time-dependent calculations are in very good agreement with each other,  as can also be seen in Fig.~\ref{fig:delay_mf_Epp_1d} where lineouts of the delays for $\theta=25^\circ,45^\circ$, and 60$^\circ$ are shown.
Figure~\ref{fig:delay_mf_Ep_1d} shows such delay lineouts for the $B\,^2\!E^{'}$ target state. The streaking delay, although somewhat affected by the 1.6-eV bandwidth of the high frequency pulse, also agrees well with the RABBIT delays.

As in the laboratory frame results, the one- and two-photon delays are in good agreement for emission angles close to the polarization axis above 5~eV (Figs.~\ref{fig:delay_mf_Epp_1d}a and~\ref{fig:delay_mf_Ep_1d}a) but the delays start to quantitatively differ at larger angles even though the shape is still similar (panels b and c in Figs.~\ref{fig:delay_mf_Epp_1d} and~\ref{fig:delay_mf_Ep_1d}). This disagreement at large angles again does not originate from any energy dependence of the continuum-continuum phases but from the geometric and magnitude effects described for the chlorine anion in Sec.~\ref{sec:chlorine_dynamics}. Dominant resonance~I at 5.5~eV gives rise to basically isotropic delays except for angles where the one- or two-photon signal vanishes. Broader resonance~III at 23~eV is still visible but the delays are strongly angle-dependent.

As in the case of laboratory-frame measurements, if oriented molecule measurements can be made, observations using electrons ejected close to the polarization axis most accurately reflect the underlying Wigner delays and allow effectively direct access to dynamics in the molecular frame. 

\section{\label{sec:core_dynamics}Core detachment dynamics of NO\boldmath{$_3^-$}}

\begin{figure*}[t!]
\includegraphics[scale=1.0]{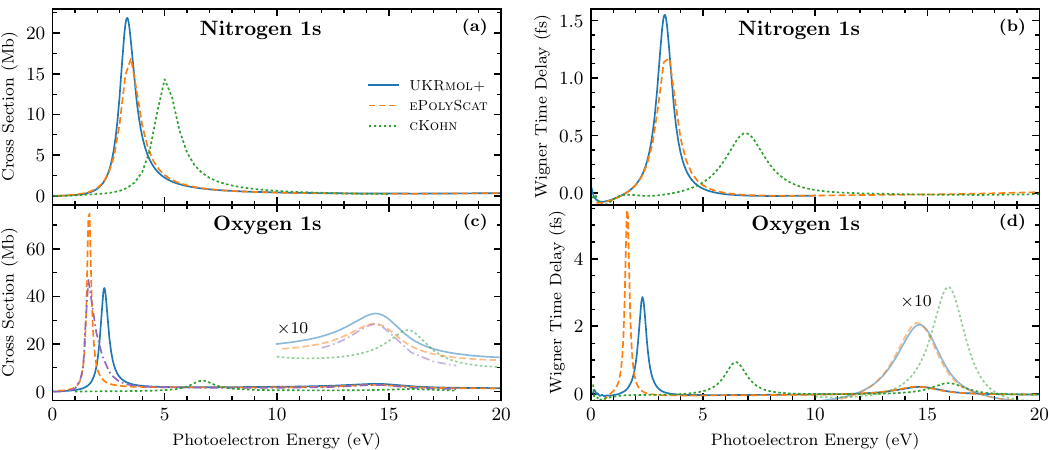}
\caption{\label{fig:core_xsec_delays}
Comparison of calculated totally-averaged integral one-photon cross sections and corresponding Wigner time delays for the photodetachment of nitrogen (panels~a and~b) and oxygen (panels~c and~d) $1s$ electrons of NO$_3^-$ using our \textsc{UKRmol+}, \textsc{ePolyScat}, and \textsc{cKohn} models. The results for the oxygen $1s$ detachment are summed over the two practically degenerate target channels. The dashed-dot purple curve for oxygen~$1s$ (panel c) is the \textsc{cKohn} calculation with the HF orbitals used in the \textsc{ePolyScat} calculation (see text). The partially transparent curves for the oxygen case above 10~eV are the results multiplied by a factor of 10.
}
\end{figure*}


\begin{figure}[]
\includegraphics[width=0.5\textwidth]{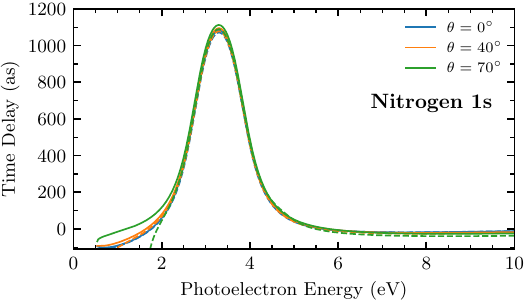}
\caption{\label{fig:N1s_delays}
One-photon (solid lines) and perturbation-theory RABBIT (dashed) time delays for the nitrogen~$1s$ photodetachment of NO$_3^-$ in the laboratory frame for electron emission polar angles $\theta=0^\circ,40^\circ$, and $70^\circ$ defined with respect to the polarization axis.
}
\vspace{1em}
\includegraphics[width=0.5\textwidth]{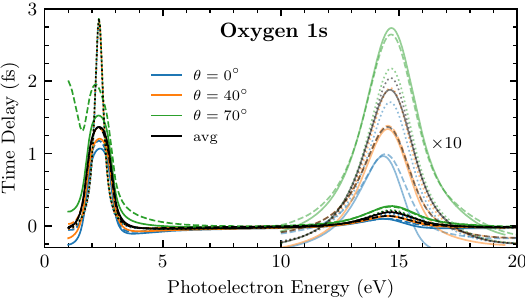}
\caption{\label{fig:O1s_delays}
Wigner (dotted), one-photon (solid lines) and perturbation-theory RABBIT (dashed) time delays for the oxygen~$1s$ photodetachment of NO$_3^-$ in the laboratory frame for electron emission polar angles $\theta=0^\circ,40^\circ$, and $70^\circ$ defined with respect to the polarization axis and also averaged delays over the $\theta$ angle (labeled as \emph{avg}). The data for energies above 10~eV are also shown multiplied by a factor of 10 drawn with partially transparent lines.
}
\end{figure}

Although the valence photodetachment dynamics and RABBIT measurements we discussed in Sections~\ref{sec:one_photon_dynamics}, \ref{sec:two_photon_dynamics_LF}, and~\ref{sec:two_photon_dynamics_MF} naturally extend traditional photodetachment experiments using UV radiation~\cite{weaver_examination_1991, babin_high-resolution_2020}, core photodetachment from an anion using X-rays is arguably a more promising type of experiment for measuring time delays. Here, we analyze the detachment dynamics of nitrogen and oxygen $1s$ electrons of NO$_3^-$ to suggest what resonances and features might be seen, and also simulate associated possible RABBIT measurements.  We will turn below to the advantages of such experiments in Sec.~\ref{sec:discussion}.

In the nitrogen case, the detachment of the $1s$ electron produces a totally-symmetric $^2\!A_1^{'}$ core-hole state of NO$_3$.  We performed several single-channel calculations with the \textsc{UKRmol+}, \textsc{ePolyScat}, and \textsc{cKohn} approaches, all using the same basis of molecular orbitals  to represent the bound and continuum electrons that were employed in the valence photodetachment calculations with these methods. The nitrogen $1s$ detachment energy was set to 401.6 eV.

In all our calculations for nitrogen $1s$ detachment, we clearly see a narrow shape resonance 4--7~eV above the detachment threshold, however, the resonance position and width significantly depend on the specific orbitals used in the calculations, as shown in Figs.~\ref{fig:core_xsec_delays}a and~\ref{fig:core_xsec_delays}b where the resulting totally-averaged one-photon cross sections and Wigner time delays are plotted.
The sensitivity of the positions and widths of  core-hole resonances to the target orbitals used in the scattering calculations is well known, as recently demonstrated for example by Guo et al.~\cite{ji_attosecond_2025}. 
The \textsc{UKRmol+} and \textsc{ePolyScat} results, which used HF orbitals give the resonance position at $\sim$3.5~eV with averaged Wigner delay of 1.2--1.6~fs. This is substantially lower than the 7-eV position of this resonance in the \textsc{cKohn} calculation, where the same resonance leads to a 500~as Wigner delay. This difference stems from the fact that the \textsc{cKohn} calculations used correlated MCSCF orbitals, which produce two optically forbidden $^2A_2$ 1s hole states {\em below} the resonance state, which is consequently pushed to higher energy. Naturally, as the energy of the resonance state lowers, the temporarily attached electron must tunnel through a wider centrifugal barrier to escape, resulting in a longer lifetime and correspondingly a narrower spectral width.
We expect the \textsc{cKohn} model to be the most realistic because correlation is ignored in the other calculations.

Using the \textsc{UKRmol+} approach, we also calculated the RABBIT delays (using IR energy of 0.517~eV again) in the laboratory frame. The one-photon dynamics is heavily dominated by single partial wave $\ell=2$, 
and thus, the RABBIT paths go essentially through only  this single intermediate partial wave.  This situation is thus  analogous to the photodetachment of atomic hydrogen anion we discussed in Sec.~\ref{sec:hydrogen_dynamics}. As a result, both the one-photon and RABBIT time delays are effectively angle-independent and identical, as can be seen in Fig.~\ref{fig:N1s_delays}. The full polar plot is shown in Fig.~9 in the SM~\cite{SM}. 

In  oxygen core photodetachment, the $1s$ electron can be detached from one of the three oxygen atoms of NO$_3^-$, leading to delocalized and effectively degenerate $^2\!A_1^{'}$ and $^2\!E^{'}$ states of NO$_3$. The energy splitting of the $^2\!A_1^{'}(^2A_1)$ and $^2\!E^{'}(^2A_1,^2B_2)$ states is only around 3~meV,  so  both states must be included in the close-coupling expansion. We performed such two-channel calculations using the \textsc{UKRmol+} and \textsc{cKohn} approaches.
In an alternative approach, Marante et al.~\cite{Marante2020} showed that such multichannel calculations can be equivalently replaced by single-channel calculations with a localized $1s$ target state, which we performed using the \textsc{ePolyScat} suite of codes for comparison. 
In the \textsc{UKRmol+} model, we again used the same orbitals and same basis for the electron in continuum as in the valence dynamics. 
With the \textsc{ePolyScat}, we computed orbitals at a slightly asymmetric geometry, with one N-O bond 0.01~\AA~ longer than the other two N-O bonds.  This geometry then belonged to the $C_{2v}$ point group, leading to the localization of one O~$1s$ orbital on the oxygen atom with the longer N-O bond, effectively decoupling it from the other two O~$1s$ orbitals.  Then using the observation in Ref.~\cite{Marante2020}, we performed a single-channel detachment calculation with the hole on the oxygen atom with the localized oxygen $1s$ orbital, with a detachment threshold set to 532 eV. This result is expected to yield a cross section which is one third of the cross section that would be obtained from the corresponding two-channel calculations where the core holes are on all three oxygen atoms and it should yield the same time delays. Additionally, we used the HF orbitals obtained from valence photodetachment with \textsc{ePolyScat} in a coupled-channel calculation with the \textsc{cKohn} method. Finally, we repeated the coupled-channel \textsc{cKohn} calculations using correlated MCSCF orbitals. 

Two shape resonances appear below 20~eV in the detachment of oxygen $1s$ electron. As in detachment of a nitrogen 1s electron, their position and width is sensitive to the orbitals used to define the target states.  Our  calculated one-photon cross sections and Wigner delays averaged over molecular orientations and emission angles are shown in Figs.~\ref{fig:core_xsec_delays}c and~\ref{fig:core_xsec_delays}d. The data are summed over the two nearly degenerate target states (three components in total).
The \textsc{UKRmol+} and \textsc{ePolyScat} models predict the higher resonance at 14.5~eV with 200~as delay. The \textsc{cKohn} model, in contrast, predicts weaker resonance features at 6.5 and 16~eV with 900 and 300~as delays, respectively. Once again, the \textsc{cKohn} model predicts that the use of correlated MCSCF orbitals and not channel coupling is the most important factor affecting the results. 

We again calculated the perturbation-theory RABBIT delays with the \textsc{UKRmol+} model using IR energy of 0.517~eV.
In contrast to nitrogen $1s$ detachment, partial waves with $\ell=0,1,2$ and $\ell=1,2,3,4$ (in different molecular symmetries) significantly contribute to the one-photon dynamics of the oxygen core case in the energy regions of the lower and higher resonance.
The interference of these partial waves results in the angular dependence of the one-photon and RABBIT delays but they follow each other rather closely. Figure~\ref{fig:O1s_delays} shows delay lineouts for $\theta=0^\circ,40^\circ$, and $70^\circ$. The polar plot is shown in Fig.~10 in the SM~\cite{SM}.
Because of the narrow resonances, the Wigner delay, also shown in Fig.~\ref{fig:O1s_delays}, notably differs from the one-photon delay due to the finite-difference approximation with $2\omega$ ($\omega=0.517$~eV) used in the one-photon delay defined by Eq.~\eqref{eq:1hv_delay}.

As in all cases in this study, the RABBIT signal effectively vanishes for electron emission direction perpendicular to the IR polarization, and $\pi/(2\omega)$ jumps appear in the RABBIT delay. 

\section{\label{sec:discussion}Discussion of experimental prospects}

In the previous sections, we showed that the photodetachment dynamics of NO$_3^-$ at electron energies up to 25~eV above the threshold is dominated by several shape resonances in both valence and nitrogen or oxygen $1s$ cases. Thanks to the absence of the long-range Coulomb potential in the photodetachment of anions, the calculated RABBIT and streaking time delays agree very well with the one-photon delays with effectively no continuum-continuum delays for electron energies above 5~eV for emission angles close to the IR polarization axis.  In other words, the long Wigner delays our calculations predict at the shape resonances could be directly measured. 

We are not aware of any experimental work measuring attosecond time delays in photodetachment of negatively charged ions. The primary challenge for such experiments is producing beams of anions with sufficient densities for attosecond experiments. 
If the experimental challenges are overcome, the nitrate anion presents a richly interesting polyatomic system to study. Its one-photon valence detachment dynamics has been experimentally studied by Neumark and collaborators~\cite{weaver_examination_1991, babin_high-resolution_2020}. However, extending such experimental studies on valence photodetachment to time delay measurements has additional challenges beyond anion production. The lower-lying resonance is located around 9~eV above the energy of the initial state, requiring in the case of RABBIT high harmonics in the 8--10~eV range, which are in principle possible with longer IR wavelengths (1.5--2.0~$\mu$m). On the other hand, 1.0--1.5-$\mu$m IR fields (1.2--0.8~eV) are in near resonance with transitions between the low-lying target states $X\,^2\!A_2^{'}$, $A\,^2\!E^{''}$, and $B\,^2\!E^{'}$ with spacings of  about 1~eV.  The probing IR field would likely induce transitions among these states. Furthermore, the dynamics of the nuclei~\cite{viel_accurate_2021, williams_simulation_2022, mahapatra_effects_2007, mukherjee_beyond_2018} may not be negligible due to the long trapping times of the lower-lying resonance that leads to time delays of 500~as.  

Such complications (except for the production of the anions) are not present in the photodetachment of nitrogen and oxygen $1s$ electrons. A tunable X-ray attosecond source with energy near the oxygen K-edge ($\sim$550~eV) could be used to detach the oxygen 1s electrons leaving them with 1--20 eV of kinetic energy. The same X-ray pulse would also produce high-energy electrons with $\sim$120~eV and effectively zero time delay from the nitrogen $1s$ detachment serving as a time reference for the delay measurement of the slow electrons affected by the two low-lying resonances in the oxygen case. Such self-referencing time delay experiments have been recently performed for photoionization of the NO molecule by Driver \textit{et al.}~\cite{driver_attosecond_2024} and for azabenzene molecules by Ji \textit{et al.}~\cite{ji_attosecond_2025} using the angular streaking method at the LCLS X-ray free-electron laser at SLAC National Accelerator Laboratory~\cite{li_co-axial_2018, li_characterizing_2018, hartmann_attosecond_2018, duris_tunable_2020, li_attosecond_2022, guo_experimental_2024}. 

After the photodetachment, the $1s$ hole will decay via the Auger-Meitner process in 4--5~fs~\cite{nicolas_lifetime_2012} and the originally neutral residual molecule becomes positively charged and suddenly the detached electron will feel the long-range Coulomb interaction. However, even if the detached electron with kinetic energy of 5~eV is temporarily detained by the shape resonance by 1~fs, it still will travel 75--100 bohr away from the molecule in 3--4~fs. Therefore, we expect that the long-range interaction with the residual ion and the post-collision interaction with the Auger-Meitner electron might have only minor effects on the detached electron and the measured time delay.  

In this work, we used relatively simple models for the NO$_3^-$ system and did not perform any time-dependent calculations for the core detachment dynamics. Motivated by experimental results, the current calculations could be extended with more accurate descriptions of the target states. Furthermore, the angular streaking process could be simulated by the time-dependent R-matrix method to also study the effect of the circular IR polarization on the direct measurability of the Wigner delay.

\section{\label{sec:conclusions}Conclusions}

We have presented results of our theoretical study of the ultrafast photodetachment dynamics of atomic hydrogen, chlorine anions, and molecular nitrate anion, where the latter molecule was our primary focus. We use the chlorine case to show in detail that even in the case of photodetachment at higher electron energies (above 5~eV) where the continuum-continuum phases are energy and partial-wave independent, the angular dependence of the Wigner and RABBIT delays are in general different when multiple intermediate partial waves contribute to the RABBIT process. 

In the main body of this study we focused on valence and core photodetachment of NO$_3^-$. Two pronounced shape resonances at energies of 5.5 and 23~eV above the detachment threshold affect the dynamics in the valence case, leading to Wigner delays up to 1~fs and 100~as, respectively. Our RABBIT and streaking simulations reproduce the Wigner delays with negligible continuum-continuum delays for electron energies above 4~eV and emission directions along the IR (and UV) polarization axis. For larger emission angles, relative to the light polarization direction, the delays are often close but not the same due to the angular effects originating from the interference of intermediate partial waves after the absorption of the first photon, as we saw in  the chlorine anion case. 

A single shape resonance appears in the detachment of the nitrogen $1s$ electron at electron energy around 7~eV with Wigner delay of 500~as. In the oxygen $1s$ detachment, two shape resonances appear at 6.5~ and 16~eV with Wigner delays of 900 and 300~as, respectively. The RABBIT delays again are close to the underlying Wigner delays.

Due to the absence of the long-range Coulomb interaction between the detached electron and residual neutral molecule, the close to one-femtosecond long Wigner delays in NO$_3^-$ would be directly measurable. The primary purpose of our work is to stimulate experimental efforts leading to measurement of attosecond (femtosecond) time delays in photodetachment of nitrate anion or other molecular anions with low-lying shape resonances.  Such experiments would effectively probe the electron-molecule scattering in real time. 
The oxygen $1s$ detachment of NO$_3^-$ with a tunable X-ray attosecond laser source using fast electrons detached from nitrogen $1s$ orbital as a time reference is a good candidate for such an experiment if technical challenges with producing sufficient amounts of anions for ultrafast experiments can be overcome.

\begin{acknowledgments}

We thank Daniel S. Slaughter and Zden\v{e}k Ma\v{s}\'{i}n for fruitful discussions.

The work at Lawrence Berkeley National Laboratory (LBNL) was supported by the Office of Science, Office of Basic Energy Sciences through the Atomic, Molecular, and Optical Sciences Program of the Division of Chemical Sciences, Geosciences, and Biosciences of the U.S. Department of Energy (DOE) under Contract No. DE-AC02-05CH11231. 
The calculations used the Lawrencium computational cluster resource provided by the IT Division at LBNL (supported by the same DOE contract as above) and also the resources of the National Energy Research Scientific Computing Center (NERSC), a DOE Office of Science User Facility. 
Some computational resources were also provided by the e-INFRA CZ project No. 90254, supported by the Ministry of Education, Youth, and Sports of the Czech Republic. 
C.S.T. was supported in part by the U.S. Department of Energy, Office of Science, Office of Workforce Development for Teachers and Scientists (WDTS) under the Berkeley Lab Undergraduate Faculty Fellowship (BLUFF) program.

\end{acknowledgments}

\appendix

\section{\label{app:lab_frame}Averaging over molecular orientations}

Here, the transformation from the molecular frame to the laboratory frame is briefly reviewed assuming a sample of randomly oriented molecules. A detailed discussion for single- and two-photon photoionization time delays is given by Baykusheva and W\"{o}rner~\cite{baykusheva2017}.

In the laboratory frame, the photoelectron direction is described by polar~$\theta$ and azimuthal~$\varphi$ angles defined with respect to a fixed polarization direction, $\hat{z}$. To obtain the single-photon cross section $\bar{\sigma}(E, \theta)$ in the laboratory frame as a function of electron energy and polar angle, we average the molecular-frame cross section given by Eq.~\eqref{eq:xsec} over molecular orientations $\hat{R}_\gamma$ using Euler angles and integrate over the azimuthal angle:
\begin{equation}
    \label{eq:xsec_lab_frame}
    \bar{\sigma}(E,\theta) = \frac{1}{8\pi^2}\int\int \sigma(E,\hat{k},\hat{R}_\gamma) \mathrm{d}\hat{R}_\gamma\mathrm{d}\varphi.
\end{equation}
For linear light polarizations, the averaged result is independent of $\varphi$, thus, the integral in the equation above can be replaced by the value for $\varphi=0$ multiplied by a factor of $2\pi$. 
The integral cross section is then obtained by integrating Eq.~\eqref{eq:xsec_lab_frame} over the polar angle:
\begin{equation}
    \bar{\sigma}(E) = \int \bar{\sigma}(E,\theta) \sin\theta \mathrm{d}\theta.
\end{equation}

To calculate the Wigner time delay, the molecular-frame delay given by Eq.~\eqref{eq:wigner_delay} is weighted by the cross section and averaged over the molecular orientations:
\begin{equation}
\begin{split}
   \bar{\tau}_W(E,\theta) = &\frac{1}{\bar{\sigma}(E,\theta)} \frac{1}{8\pi^2}\\
    & \times\int \tau_W(E,\hat{k},\hat{R}_\gamma) \sigma(E,\hat{k},\hat{R}_\gamma) \mathrm{d}\hat{R}_\gamma\mathrm{d}\varphi,
    \end{split}
\end{equation}
\begin{equation}
    \bar{\tau}_W(E) = \frac{1}{\bar{\sigma}(E)}\int \bar{\tau}_W(E,\theta)\bar{\sigma}(E,\theta) \sin\theta \mathrm{d}\theta.
\end{equation}

In the RABBIT experiments, the time delay is inferred from the measured signals with the \emph{cosine} modulations, see Eq.~\eqref{eq:rabitt_signal_cos}.
In the case of laboratory-frame measurement, the delay is determined from the measured averaged signal, and thus, 
we need to first average the perturbation-theory RABBIT interference term and extract the phase afterwards:
\begin{equation}
    \bar{\tau}_R(E,\theta) = \frac{1}{2\omega} \arg \int M^{(+)*}(\hat{R}_\gamma) M^{(-)}(\hat{R}_\gamma)\mathrm{d}\hat{R}_\gamma,
\end{equation}
\begin{equation}
    \bar{\tau}_R(E) = \frac{1}{2\omega} \arg \int\int M^{(+)*}(\hat{R}_\gamma) M^{(-)}(\hat{R}_\gamma)\mathrm{d}\hat{R}_\gamma \sin\theta \mathrm{d}\theta.
\end{equation}

\section{\label{app:threshold_delay}Behavior of scattering phaseshift at threshold}

In scattering from a spherically symmetric potential, the radial wave function for partial wave $\ell$ has the following asymptotic form:
\begin{equation}
    \psi_{k\ell}(r) \xrightarrow{r\rightarrow\infty} e^{i\delta_\ell(k)}\sin\left(kr -\frac{\pi}{2}\ell + \delta_\ell(k)\right),
\end{equation}
where $\delta_\ell$ is the scattering phaseshift depending on the electron asymptotic momentum~$k$. 
Using the effective range theory the phaseshift behavior near threshold for short-range potentials is given by~\cite{taylor1972scattering}
\begin{equation}
    \delta_\ell(k) \xrightarrow{k\rightarrow 0} -a_\ell k^{2\ell+1},
\end{equation}
where $a_\ell$ is the scattering length.

The Wigner delay near threshold is then given by
\begin{equation}
    \tau_\ell=\frac{\mathrm{d}\delta_\ell(k)}{\mathrm{d}E} = \frac{1}{k}\frac{\mathrm{d}\delta_\ell(k)}{\mathrm{d}k} \xrightarrow{k\rightarrow 0} -(2\ell+1) a_\ell k^{2\ell-1}.
\end{equation}
The delay diverges for $\ell=0$ and goes to zero for $\ell\geq 1$.

However, for polarization potentials $-\alpha/(2r^4)$ with polarizability $\alpha$, modifying the effective range formula for $p$~wave gives~\cite{omalley1962}
\begin{equation}
    \tan \frac{\delta_1}{k^2} = \frac{\pi\alpha}{15} - A_1 k,
\end{equation}
where $A_1\approx 1.6$ in atomic units for singlet electron-hydrogen atom scattering. As a result, the $p$-wave time delay limits to a constant at threshold determined by the atom's polarizability:
\begin{equation}
    \tau_1 \xrightarrow{k\rightarrow 0} \frac{2\pi\alpha}{15}.
\end{equation}
The polarizability of hydrogen atom is $\alpha=4.5$ in atomic units and the delay limits to 45.6~as at $k=0$.
The polarization potential doesn't change the divergent behavior of the time delay for an $s$~wave~\cite{omalley1962}, and that behavior is seen in valence photodetachment of chlorine atom where the electron originates from the $2p$ orbital.

\section{\label{app:Born_RABBIT_amplitudes} Born approximation for the RABBIT amplitudes}

The partial wave amplitudes for the two-photon RABBIT transitions for photodetachment of H$^-$ (described by a single configuration), can be found by expanding Eq.(\ref{eq:two_photon_element}) in partial waves and reducing the expression to a single active electron: 
\begin{equation}
\begin{split}
&M^\pm_{l_f} = \\
& \frac{-i}{\hbar}\,i^{l_f} \sqrt{\frac{2}{\pi}} \\
&\times \Big\langle \frac{ \hat{j}_{l_f}(k_f\,r)}{k_r\,r} Y_{l_f,m_f}\Big\vert   \hat{\varepsilon}\cdot \vec{r}  \, \,  G^+(E_0 + \hbar\Omega_\pm)  \, \, \hat{\varepsilon}\cdot \vec{r} \, \Big\vert \frac{\varphi_0}{r} \, Y_{0,0} \Big\rangle \\
=&  \frac{2i^{l_f-1} }{k_f }  \sqrt{\frac{2}{\pi}}  \sum_{l} \int_0^\infty \int_0^\infty \hat{j}_{l_f}(k_f\,r) rG_l^+(r,r') r' \varphi_0(r') \, dr \, dr' \\
&\times \frac{4 \pi}{3} \int Y^*_{l_f,0} Y_{1,0}Y_{l,0} \, d\Omega\, \int Y^*_{l,0}Y_{1,0} Y_{0,0} \, d\Omega, \\
\end{split}
\label{eq:Mpartialwave}
\end{equation}
where $\hat{j}_l$ denotes the regular Riccati-Bessel function, $\varphi_0(r)$ is the radial function of the H$^-$ 1s orbital, and we have specialized to z-polarization and used atomic units in the second line. The angular integrations require $l = 1$ in this simple case.  The energy of the RABBIT sideband is 
\begin{equation}
\frac{ k_f^2}{2} = E_0 + \Omega_\pm \pm \omega 
\end{equation}
where $E_0$ is the energy of the initial state, and so the Green's function, is evaluated at $k=\sqrt{k_f^2 \mp 2\omega}$.

At high enough kinetic energies, which turn out to be just a few eV in the case of H$^-$, especially in the static-exchange approximation, we can replace $G_l$ by the free-particle Green's function~\cite{taylor1972scattering}, 
\begin{equation}
\begin{split}
G_l^+(r,r') 
= \frac{2}{\pi} \int_{-\infty}^{+\infty} \frac{ \hat{j}_l(p\,r) \hat{j}_l(p\,r') }{ k^2 - p^2   +  i \epsilon } dp \, 
\end{split}
\label{eq:G0spect}
\end{equation}
and make the simple approximation  for the initial state, $\varphi_0(r) = \zeta ^{3/2} 2 r e^{-\zeta r}$.  To evaluate the radial integrals in Eq.(\ref{eq:Mpartialwave}) we can use  the following identities for the free-free dipole integrals,
\begin{equation}
\begin{split}
\frac{2}{\pi} \int_{-\infty}^\infty \hat{j}_{l-1}(k_f r) r \hat{j}_l(pr) dr &= \frac{l}{k_f} \delta(k_f-p) + \delta'(k_f-p), \\
\frac{2}{\pi} \int_{-\infty}^\infty \hat{j}_{l+1}(k_f r) r \hat{j}_l(pr) dr &= \frac{l+1}{k_f} \delta(k_f-p)- \delta'(k_f-p), \\
\frac{2}{\pi} \int_{-\infty}^\infty \hat{j}_l(k_f r) \hat{j}_l(p r) dr &= \delta(k_f-p), 
\end{split}
\label{eq:bessel_identities}
\end{equation}
whose delta functions and their derivatives simplify the integral over $p$ from Eq.(\ref{eq:G0spect}) in the Green's function.  The full Born RABBIT amplitudes then become
\begin{equation}
\begin{split}
M^-_{p\to s} &= -\frac{16 i \sqrt{\frac{2}{\pi }} \zeta ^{5/2} \left(-3 \zeta ^2 \omega+k_f^4+k_f^2 \left(\zeta ^2+3 \omega\right)\right)}{3 \omega^2 \left(\zeta ^2+k_f^2\right)^4}  \\
M^-_{p\to d} &= -\frac{32 i \sqrt{\frac{2}{5 \pi }} \zeta ^{5/2} k_f^2 \left(\zeta ^2+k_f^2+6 \omega\right)}{3 \omega^2 \left(\zeta ^2+k_f^2\right)^4}  \\
M^+_{p\to s} &= -\frac{16 i \sqrt{\frac{2}{\pi }} \zeta ^{5/2} \left(3 \zeta ^2 \omega+k_f^4+k_f^2 \left(\zeta ^2-3 \omega\right)\right)}{3 \omega^2 \left(\zeta ^2+k_f^2\right)^4}  \\
M^+_{p\to d} &= -\frac{32 i \sqrt{\frac{2}{5 \pi }} \zeta ^{5/2} k_f^2 \left(\zeta ^2+k_f^2-6 \omega\right)}{3 \omega^2 \left(\zeta ^2+k_f^2\right)^4}
\end{split}
\end{equation}
These amplitudes have a fixed complex phase factor of $\exp{(-i \pi/2)}$ at energies of a few eV, and thus, they have no time delay but are not strictly factorable into two single-photon amplitudes.

However, if we replace the partial-wave, free-particle Green's function [Eq.~\eqref{eq:G0spect}] by its asymptotic form 
\begin{equation}
G_l^+(r,r') \xrightarrow{r\to\infty} -\frac{1}{k} \hat{h}^+_1(k\,r) \hat{j}_1(k\,r'), 
\end{equation}
where $h_1^+$ is the Ricatti-Hankel function with $l=1$,
the resulting approximate Born amplitudes are strictly factorable, like the approximation in Eq.(\ref{eq:Mfactors}), and become in this case 
\begin{equation}
\begin{split}
M^\pm_{l_f} \approx & \,  M^1 (k_\pm)  \, C_{l_f} (k_\pm)  \\
C_{l_f} (k_\pm) = &  \frac{-i}{\hbar}\,i^{l_f} \sqrt{\frac{2}{\pi}}  \int_0^\infty \hat{j}_{1_f}(k_f\,r) \, r \, \hat{h}^+_1(k_\pm\,r) dr \\
&  \times \sqrt{\frac{4 \pi}{3}}  \int Y^*_{l_f,0} Y_{1,0}Y_{l,0} \,  d\Omega  \\
 M^1 (k_\pm) =  & \frac{1}{\sqrt{3}} \int_0^\infty  \hat{j}_1(k_\pm\,r) \, r \, \varphi_0(r) dr \\
\end{split}
\end{equation}
with $k_\pm^2/2 = k_f^2/2 \mp \omega$.   The factors of these approximate, asymptotic Born RABBIT amplitudes are
\begin{equation}
\begin{split}
C_{p \to s} (k_\pm) &= -\frac{i \sqrt{\frac{2}{3 \pi }} \left(k_f^2 \mp 3 \omega\right)}{\omega^2 \left(k_f^2 \mp 2 \omega\right)} \\
C_{p \to d} (k_\pm) &= -\frac{2 i \sqrt{\frac{2}{15 \pi }} k_f^2}{\omega^2 \left(k_f^2 \mp 2 \omega\right)} \\
 M^1 (k_\pm)   & = \frac{16 \zeta ^{5/2} \left(k_f^2 \mp 2 \omega\right)}{\sqrt{3} \left(\zeta ^2+k_f^2  \mp  2 \omega\right)^3} \\
\end{split}
\label{eq:BornAsymp}
\end{equation}

\begin{figure}[]
\includegraphics[scale=1]{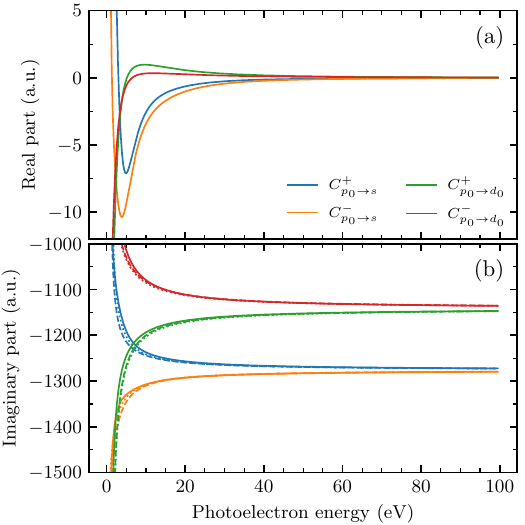}
\caption{\label{fig:hminus_born_cc_amplitudes}
Comparison of analytical Born and computed two-photon amplitudes from the R-matrix calculations in the static exchange approximation to the photodetachment of hydrogen anion. Panel a: Real part of the amplitudes from R-matrix calculations divided by the single-photon UV absorption amplitudes.  Panel b:  R-matrix  amplitudes divided by UV absorption amplitudes (solid),  full Born amplitudes divided by Born UV absorption amplitudes (dashed), Born continuum-continuum amplitudes from Eq.~\eqref{eq:BornAsymp} (dotted).  $\zeta = 1$ and $\lambda_\mathrm{IR} = 2.4$~$\mu$m were used for this comparison.  }
\end{figure}

Remarkably,  these expressions closely reproduce the behavior for photoelectron energies above a few eV of the RABBIT amplitudes calculated for H$^-$ using the R-matrix method that produce the delays in Fig. \ref{fig:delays_hydrogen_1d}.  That comparison is shown in Fig. \ref{fig:hminus_born_cc_amplitudes}.  The full Born expressions closely match the factorable asymptotic expressions in Eq. (\ref{eq:BornAsymp}) above a few eV.  Moreover, the RABBIT signal over the range of energies in that figure is almost exactly proportional to $\cos^4(\theta)$ and is thus effectively zero for photoejection perpendicular to the polarization axis. 

\section{\label{app:rabbitt_hydrogen_derivation}RABBIT angular dependence for photodetachment of hydrogen anion}

In this appendix, we show in detail that the RABBIT time delay for photodetachment of atomic hydrogen anion is angle independent if the continuum-continuum phases are independent of the partial waves. 
The two-photon RABBIT angular distribution, as defined in Eq.~\eqref{eq:2hv_spectrum_hm}, is given by
\begin{align}
    \label{eq:2hv_spectrum_hm_app}P_2(E,\theta,\tau) &= \vert A_0(E,\tau) Y_{00}(\theta) + A_2(E,\tau) Y_{20}(\theta)\vert^2,\\
A_0(E,\tau) &= M^+_{p \to s}\, e^{i\omega\tau} + M^-_{p \to s}\, e^{-i\omega\tau}, \\
A_2(E,\tau) &= M^+_{p \to d}\, e^{i\omega\tau} + M^-_{p \to d}\, e^{-i\omega\tau}.
\end{align}
The square in Eq.~\eqref{eq:2hv_spectrum_hm_app} gives the sum of two diagonal terms $\vert A_0\vert^2$ and $\vert A_2\vert^2$ proportional to $\vert Y_{00}\vert^2$ and $\vert Y_{20}\vert^2$ and the cross term $2\mathrm{Re}\,(A_0^* A_2)Y_{00}Y_{20}$.
To obtain the RABBIT delay, we need to investigate the terms oscillating with the time delay~$\tau$ between the high-frequency and IR fields.

We begin with the basic assumption of RABBIT that the phases of the two-photon matrix elements can be divided into the one-photon phase $\eta_p^\pm$ evaluated at the harmonic energies $\eta^\pm_p=\eta_p(E\mp\omega)$ and continuum-continuum phases $\varphi^\pm_{s/d}$:
\begin{align}
  M^\pm_{p \to s} &= |M^\pm_{p \to s}|\, e^{i(\eta_p^\pm + \varphi_s^\pm)}, \\
  M^\pm_{p \to d} &= |M^\pm_{p \to d}|\, e^{i(\eta_p^\pm + \varphi_d^\pm)}, \\
\end{align}
Then, the $\tau$-dependent part $\vert\tilde{A}_0\vert^2$ of the $\vert A_0\vert^2$ term is
\begin{align}
\vert\tilde{A}_0\vert^2 &= 
  M^+_{p\to s}\, M^{-*}_{p\to s}\, e^{+2i\omega\tau}
  + M^{+*}_{p\to s}\, M^-_{p\to s}\, e^{-2i\omega\tau} \\
  & = 2\,|M^+_{p\to s}|\,|M^-_{p\to s}|\,
  \cos(2\omega\tau + \Delta\Phi_0),
\end{align}
where 
\begin{equation}
 \Delta\Phi_0 =  \eta_p^- - \eta_p^+ + \varphi_s^- - \varphi_s^+.
\end{equation}
Analogously for the $\vert A_2\vert^2$ term, we get cosine oscillations $\cos(2\omega\tau + \Delta\Phi_2)$ with
$\Delta\Phi_2 = \eta_p^- - \eta_p^+ + \varphi_d^- - \varphi_d^+$. Finally, the $A_0^*A_2$ term leads to two cosine functions
\begin{equation}
  2\,|C_1|\,\cos(2\omega\tau + \Delta\Phi_{02a})
  + 2\,|C_2|\,\cos(2\omega\tau + \Delta\Phi_{02b}),
\end{equation}
where
\begin{align}
  C_1 &= M^{-*}_{p\to s}\, M^+_{p\to d},\\
  C_2 &= M^{+*}_{p\to s}\, M^-_{p\to d},\\
  \Delta\Phi_{02a} &= \eta_p^- - \eta_p^+ + \varphi_d^- - \varphi_s^+, \\
  \Delta\Phi_{02b} &= \eta_p^- - \eta_p^+ + \varphi_s^- - \varphi_d^+. 
\end{align}

In total, the $\tau$-dependent part of the RABBIT spectrum $P(E,\theta,\tau)$ from Eq.~\eqref{eq:2hv_spectrum_hm_app} is
\begin{align}
\label{eq:spectrum_osc_hm}
  P_2^\mathrm{osc}(E,\theta, \tau)
  &= 2\,|M^+_{p\to s}|\,|M^-_{p\to s}|\, Y_{00}^2(\theta)\,
     \cos(2\omega\tau + \Delta\Phi_0)
     \nonumber \\
  &+ 2\,|M^+_{p\to d}|\,|M^-_{p\to d}|\, Y_{20}^2(\theta)\,
     \cos(2\omega\tau + \Delta\Phi_2)
     \nonumber \\
  &+ 2\,|C_1|\, Y_{00}(\theta)\, Y_{20}(\theta)\,
     \cos(2\omega\tau + \Delta\Phi_{02a})
     \nonumber \\
  &+ 2\,|C_2|\, Y_{00}(\theta)\, Y_{20}(\theta)\,
     \cos(2\omega\tau + \Delta\Phi_{02b}).
\end{align}

In general, the sum of such cosine functions can be written as one cosine function: 
\begin{equation}
    \sum_i c_i \cos(x + \delta\varphi_i) = C\cos(x+\Delta\Phi)
\end{equation}
with $C=\sqrt{a^2+b^2}$, $\Delta\Phi=\arctan(b/a)$, where $a=\sum_i c_i\cos\delta\varphi_i$ and $b=\sum_i c_i\sin\delta\varphi_i$.
Since the amplitudes of individual cosine functions in Eq.~\eqref{eq:spectrum_osc_hm} are angle-dependent,
the resulting RABBIT phase $\Delta\Phi$ and time delay $\Delta\Phi/(2\omega)$ would be angle dependent for general continuum-continuum phases $\varphi^\pm_{\ell_f}$. At energies, where the Born approximation begins to apply in photodetachment, we show in Appendix \ref{app:Born_RABBIT_amplitudes}  that $\varphi^+_{\ell_f}=\varphi^-_{\ell_f}$ and
$\varphi^\pm_{\ell_f}=\varphi^\pm_{\ell_f^{'}}$, and thus, the individual cosine phase offsets reduce to the one-photon phase offset $\delta\varphi_1$: $\Delta\Phi_0=\Delta\Phi_2=\Delta\Phi_{02a}=\Delta\Phi_{02b}=\eta^-_p-\eta^+_p=\delta\varphi_1$.
Then, the angular dependence of the RABBIT spectrum can be factored out
\begin{equation}
  P_2^{\mathrm{osc}}(E,\theta, \tau)
  = G(\theta) \cos(2\omega\tau + \delta\varphi_1)
\end{equation}
and the RABBIT delay is angle-independent and equal to the one-photon delay,
which is demonstrated  in the static-exchange approximation to the photodetachment of atomic hydrogen anion above 5~eV in  Fig.~\ref{fig:delays_hydrogen_1d}.  In the case of hydrogen anion photodetachment, the angular dependence of the delay measured with 
RABBIT is the same as that of the Wigner delay, fundamentally because only one intermediate partial wave contributes.


\bibliography{references_photodetachment}

\end{document}